\documentclass[reprint,superscriptaddress,aps]{revtex4-2}
\usepackage{graphicx,subfigure, amsmath, amssymb, mathtools, enumitem}

\makeatletter
\providecommand*{\input@path}{}
\g@addto@macro\input@path{{sections/}}% append
\makeatother

\graphicspath{{Figure/}{../Figure/}}

\begin{document}

\title{Randomized-gauge test for machine learning of Ising model order parameter}

\author{Tomoyuki Morishita}
\affiliation{Department of Physics, The University of Tokyo, Tokyo 113-0033, Japan}

\author{Synge Todo}
\affiliation{Department of Physics, The University of Tokyo, Tokyo 113-0033, Japan}
\affiliation{Institute for Physics of Intelligence, The University of Tokyo, Tokyo 113-0033, Japan}
\affiliation{Institute for Solid State Physics, The University of Tokyo, Kashiwa 277-8581, Japan}

\date{\today}

\begin{abstract}
  Recently, machine learning has been applied successfully for identifying phases and phase transitions of the Ising models.
The continuous phase transition is characterized by spontaneous symmetry breaking, which can not be detected in general from a single spin configuration.
To investigate if neural networks can extract correlations among spin snapshots, we propose a new test using the random-gauge Ising model.
We show that neural networks can extract the order parameter or the energy of the random-gauge model as in the ferromagnetic case.
We also discuss how and where the information of random gauge is encoded in neural networks and attempt to reconstruct the gauge from the neural network parameters.
We find that the fully connected network encodes the effect of random gauge to its weights naturally.
In contrast, the convolutional network copes with the randomness by assigning different network parts to local gauge patterns.
This observation indicates that although the latter demonstrates higher performance than the former for the present randomized-gauge test, the former is more effective and suitable for dealing with models with spatial randomness.

\end{abstract}

\maketitle

\section{Introduction}
\label{sec:Introduction}

Machine learning~\cite{Hinton2006, LeCun2015} has become one of the most powerful tools for data analysis and has already been used in a wide range of natural and social sciences~\cite{Jordan2015}, such as image classification~\cite{Bishop2006} and natural language processing~\cite{Hinton2012}.
Recently, we have seen that machine learning has also been applied in various branches of physics~\cite{Carleo2019}.
Especially in the fields of condensed matter and statistical physics, neural network (NN) is employed for identifying phases of matter and phase transitions~\cite{Broecker2017a, PhysRevB.96.205146, PhysRevB.96.144432, PhysRevLett.120.176401, PhysRevLett.120.066401, Beach2018}.
The ferromagnetic Ising model is one of the simplest statistical models that exhibit a continuous phase transition, and thus there are many previous studies of the model using NN~\cite{VanNieuwenburg2017, Wetzel2017a, Wetzel2017, Carrasquilla2017, Morningstar2018, Suchsland2018, Iso2018, Kim2018, Kashiwa2019}.
For instance, Kashiwa, Kikuchi, and Tomiya~\cite{Kashiwa2019} demonstrated that after NN is trained by feeding Ising spin configurations with temperature labels, the magnetization or the energy of the system is automatically extracted at an intermediate NN state (or output) depending on the type of NN.

These previous studies appear to indicate the usefulness of NN applying to the Ising model.
However, one may think that NN implicitly takes advantage of the fact that the uniform magnetization is the order parameter or the nearest-neighbor correlation is the energy in the ferromagnetic Ising model;
In other words, the success may be attributed not to the extraordinary ability of NN but the apparent connection between the spin configuration and the relevant physical quantities in the model.
In general cases, one can not detect the spontaneous symmetry breaking associated with the continuous phase transition from a single spin snapshot.
To confirm whether NN can extract a physical essence of the phase transition, we need to subject NN to a more sophisticated test, in which detecting correlations among spin snapshots is essential than the spatial correlation within a single spin configuration.

In the present paper, we propose a new test using the random-gauge Ising model, a model physically equivalent to the ferromagnetic Ising model but its magnetization is shuffled by a randomly chosen gauge, and examine whether NN can perform the same task for the random-gauge Ising model as for the ferromagnetic one.
We choose Ref.~\cite{Kashiwa2019} as the task to be carried out.
By feeding Ising spin configurations perturbed by the random gauge with temperature labels, we verify that NN can extract the order parameter and the energy correctly.
We also discuss how and where the information of random gauge is encoded in NN and attempt to reconstruct the gauge from the NN parameters obtained through learning.

This paper is organized as follows.
In Sec.~\ref{sec:Models}, we introduce the models we study: the ferromagnetic Ising model and the random-gauge Ising model.
Following Sec.~\ref{sec:Method} explains the dataset and the NN architecture adopted in the present study.
Then, in Sec.~\ref{sec:Results}, we compare the performance of NN for the two different models and analyze the intermediate outputs and parameters of NN to discuss how and where the gauge information is encoded in NN.
In Sec.~\ref{sec:Gauge}, to demonstrate that NN learns the gauge correctly, we attempt to reconstruct the gauge from the learned NN parameters.
In Sec.~\ref{sec:Mult}, we examine what happens to NN learning if spin configurations in the dataset are shuffled by two different gauge patterns.
Finally, we give a conclusion in Sec.~\ref{sec:Conclusion}.

\section{Models}
\label{sec:Models}

We consider two different models in the present paper: the ferromagnetic Ising model and the random-gauge Ising model.
The former is defined by the Hamiltonian:
\begin{equation}\label{eq:Ferro-Ham}
  \mathcal{H}=-\sum_{\langle r,r' \rangle}\sigma _r\sigma _{r'},
\end{equation}
where $\sigma _{r}=\pm 1$ denotes the Ising spin variable at site $r$.
We assume the two-dimensional square lattice with periodic boundary conditions and the summation in Eq.~\eqref{eq:Ferro-Ham} is taken over all nearest-neighbor pairs.
The system size is $N = L^2$.
We denote the spin configuration by an $L \times L$ matrix $S$, where $S_{i,j}$ represents the spin variable at $r = (i,j)$ on the square lattice and
\begin{equation}\label{eq:PBC}
  S_{i+aL,j+bL}=S_{i,j} \quad (a,b\in \mathbb{Z})
\end{equation}
due to the periodic boundary conditions.
The inverse temperature is denoted by $\beta \equiv 1/k_\mathrm{B} T$, where $k_\mathrm{B}$ is the Boltzmann constant and $T$ is the temperature.

The second model is the random-gauge Ising model, described by the Hamiltonian:
\begin{equation}\label{eq:RG-Ham}
  \mathcal{H}=-\sum_{\langle r,r' \rangle}\tau _r\tau _{r'}\sigma _r\sigma _{r'},
\end{equation}
where $\tau _r$ denotes the gauge for site $r$. The gauge for each site is independently and randomly quenched as $+1$ or $-1$ with probability $\frac{1}{2}$.
We also represent $\{\tau_r\}$ by an $L \times L$ matrix $\Gamma$.

The above two models~\eqref{eq:Ferro-Ham} and \eqref{eq:RG-Ham} are related
with each other through the gauge transformation:
\begin{equation}\label{eq:gauge-trans-1}
  \sigma_r^\mathrm{R} = \tau_r \sigma_r^\mathrm{F} \quad \text{for $^\forall r$,}
\end{equation}
where $\sigma_r^\mathrm{F}$ ($\sigma_r^\mathrm{R}$) denotes Ising spins of the ferromagnetic (random-gauge) model. The relation~\eqref{eq:gauge-trans-1} can be expressed in the matrix form as
\begin{equation}\label{eq:RG-Hadamard}
  S^\mathrm{R}=\Gamma \odot S^\mathrm{F},
\end{equation}
where the symbol $\odot$ represents the entrywise product:
\begin{equation}\label{eq:RG-Hadamard-Def}
  (A\odot B)_{i,j}\equiv A_{i,j}B_{i,j} \quad \text{for $i,j=1,\ldots,L$.}
\end{equation}

\begin{figure}[tbp]
  \begin{center}
    \includegraphics[width=.8\hsize]{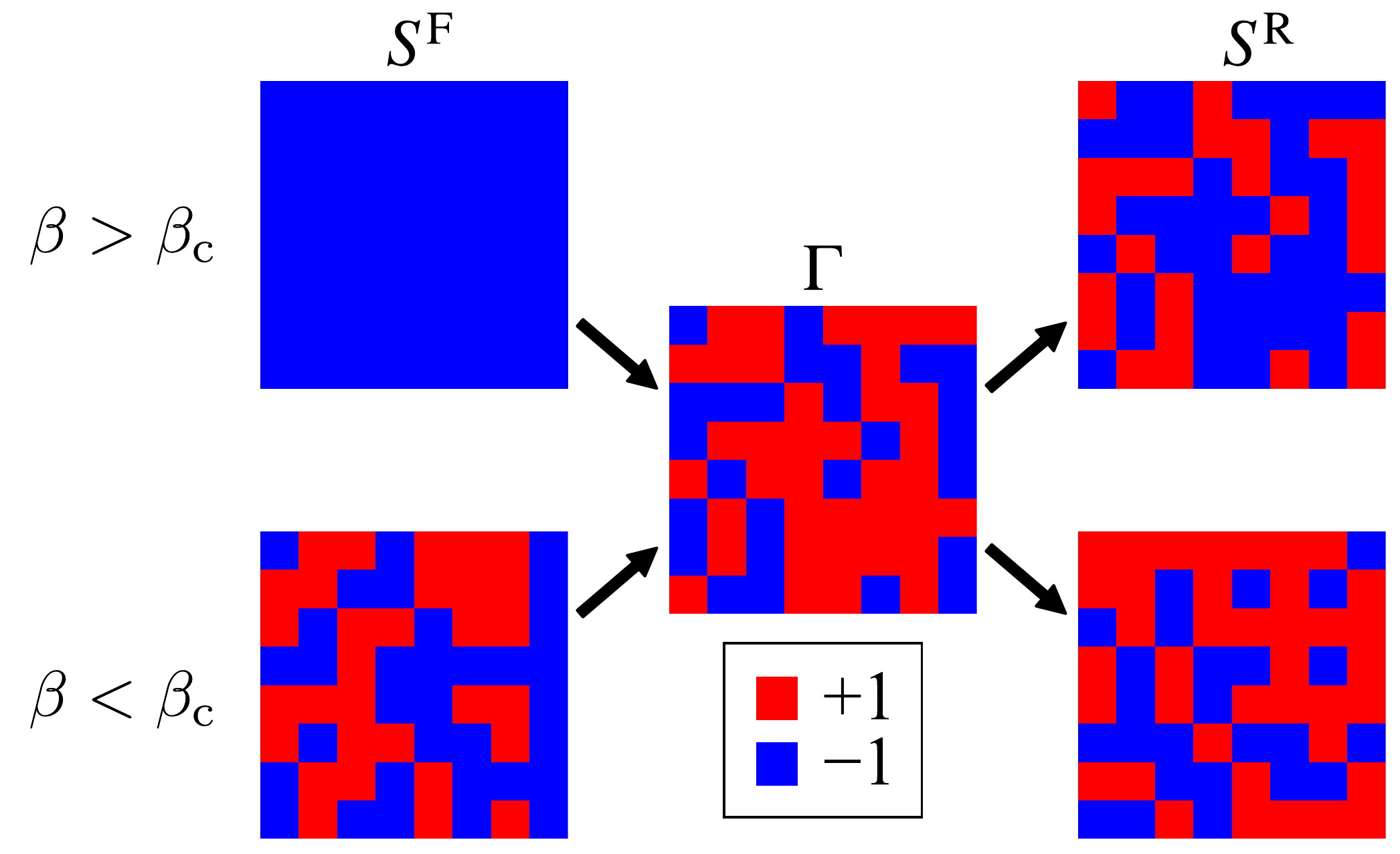}%
  \end{center}
  \caption{\label{fig:Conf}(Color online).
    Typical spin configurations of the ferromagnetic (left) and random-gauge (right) Ising models in the low/high-temperature phases.
    The red (blue) color denotes the up (down) spin.
    Spin configurations of both models are related by the gauge transformation denoted by $\Gamma$.
    Note that the spin configuration of the random-gauge model looks `random' even in the ordered phase ($\beta > \beta_\mathrm{c}$).
  }
\end{figure}

In the thermodynamic limit, $L \rightarrow \infty$, the ferromagnetic Ising model~\eqref{eq:Ferro-Ham} exhibits a continuous phase transition at $\beta = \beta_\mathrm{c} = \frac{1}{2} \log (\sqrt{2}+1) \approx 0.4407$~\cite{Kramers1941,Onsager1944}.
In the ordered phase ($\beta > \beta_\mathrm{c}$), the expectation value of the uniform magnetization:
\begin{equation}\label{eq:RG-Magnet}
  m = \frac{1}{L^2} \sum _r \sigma _r
\end{equation}
becomes finite.
The random-gauge Ising model~\eqref{eq:RG-Ham} also exhibits a phase transition at the same critical temperature $\beta = \beta_\mathrm{c}$, as it is precisely related to the ferromagnetic model via the gauge transformation~\eqref{eq:RG-Hadamard}.
It should be noted, however, that the uniform magnetization~\eqref{eq:RG-Magnet} does no longer act as an order parameter (see Fig.~\ref{fig:Conf}).
The order parameter of the random-gauge model is given by the generalized magnetization:
\begin{equation}\label{eq:generalized-magnetization}
  m^\mathrm{R} = \frac{1}{L^2} \sum _r \tau_r \sigma _r.
\end{equation}

One can rewrite the Hamiltonian of the random-gauge model~\eqref{eq:RG-Ham} as
\begin{equation}\label{eq:RG-Ham-J}
  \mathcal{H}=-\sum_{\langle r,r' \rangle}J_{r,r'}\sigma _r\sigma _{r'}
\end{equation}
by introducing $J_{r,r'} \equiv \tau_r \tau_{r'}$.
A product of $J_{r,r'}$ along any closed loop on the square lattice is equal to +1; The present random-gauge Ising model can be regarded as the Villain model with the `even rule'~\cite{Villain1977}.
We denote $\{J_{r,r'}\}$ by $\Theta$.
Note that $\Gamma$ can be determined from $\Theta$ besides the overall sign.
Considering the $\mathbb{Z}_2$ symmetry of the Ising model, we identify $\Theta$ with $\Gamma$ and use the term `gauge' for both of $\Gamma$ and $\Theta$.

\section{Dataset and Neural Network Architecture}
\label{sec:Method}

\subsection{Dataset}
\label{sec:Method-Dataset}

We prepared the input dataset for NN using the Monte Carlo technique with the Swendsen-Wang cluster algorithm~\cite{Swendsen1987}.
The system size is $N = L \times L = 64^2$.
The spin configurations in the dataset were sampled every $2^{10}$ Monte Carlo steps after discarding the first $2^{16}$ steps for thermalization.
The Monte Carlo simulations were performed at nine different inverse temperatures, $\beta =0.05, 0.10, 0.15, \ldots, 0.45$.
For each model, we sampled 2,000 (400) configurations for the training (test) dataset at each temperature, and thus the total size of the dataset is 18,000 (3,600).

Note that most of configurations in the dataset are in the high-temperature phase ($\beta < \beta_\mathrm{c}$).
Unlike the numerical experiments in Ref.~\cite{Kashiwa2019}, we excluded low-temperature configurations since at low temperatures, the samples are almost identical to the gauge itself, and it makes training and inference trivial.
Due to finite-size thermal fluctuations, the theoretical upper bound of classification accuracy is 99.2\% in the present case, $N = L \times L = 64^2$ and $\beta =0.05, 0.10, 0.15, \ldots, 0.45$.
We evaluated the upper bound in the same way as in Ref.~\cite{Kashiwa2019}, though we used the exact finite-size free energy~\cite{Kastening2001, Exact} instead of that in the thermodynamic limit.

\subsection{Neural network architecture}
\label{sec:Method-Archi}

We implemented two types of NNs, the fully connected (dense) NN and the convolutional NN~\cite{Goodfellow-et-al-2016}, using Keras~\cite{Chollet2015} with TensorFlow~\cite{Abadi2016} as a backend.
Both NNs receive spin configuration $S$ as an input and output a nine-dimensional vector $\mathbf{y}$.
As the training label, the one-hot vector representation corresponding to the nine temperature classes is used; NN weights are optimized so that NN learns the temperature class to which the input configuration belongs.
As the loss function, we used cross-entropy. We introduced the $L_2$-regularization to the NN weights to avoid overfitting.
The optimization of the NN weights is performed using the Adam optimizer~\cite{Kingma2014}.

To simplify the following discussion, we introduce some shorthand matrix notations:
\begin{itemize}
\item Sum of all matrix elements:
  \begin{equation}
    \mathrm{sum}(A)\equiv \sum_{i,j}A_{i,j}.
  \end{equation}
\item Submatrix of size $2 \times 2$:
  \begin{equation}
    \tilde{A}_{(i,j)}\equiv
    \begin{pmatrix}
      A_{i,j} & A_{i,j+1} \\
      A_{i+1,j} & A_{i+1,j+1}
    \end{pmatrix}.
  \end{equation}
  Note that the periodic boundary condition is considered appropriately, e.g., 
  \begin{equation}
    \tilde{A}_{(L,L)}\equiv
    \begin{pmatrix}
      A_{L,L} & A_{L,1} \\
      A_{1,L} & A_{1,1}
    \end{pmatrix}.
  \end{equation}
\item Entrywise sign function:
  \begin{equation}
    \left(\mathrm{sgn}(A) \right)_{i,j} \equiv
    \mathrm{sgn}(A_{i,j}).
  \end{equation}
\end{itemize}
In addition, if two matrices $A$ and $B$ satisfy
\begin{equation}
  \mathrm{sgn}(A)=\mathrm{sgn}(B)\quad \mathrm{or}\quad \mathrm{sgn}(A)=\mathrm{sgn}(-B),
\end{equation}
we say that {\em $A$ and $B$ have the same pattern}.

Note that learning of NN is affected by the gauge $\Gamma$ and the initial condition for NN weights.
We thus performed learning 1,000 times with different gauge patterns and initial conditions and took an average for evaluating the classification accuracy.

\subsubsection{Fully connected neural network}
\label{sec:Method-Archi-FNN}

\begin{figure*}
  \begin{center}
  \subfigure[Layer structure]{%
    \includegraphics[clip, width=0.4\columnwidth]{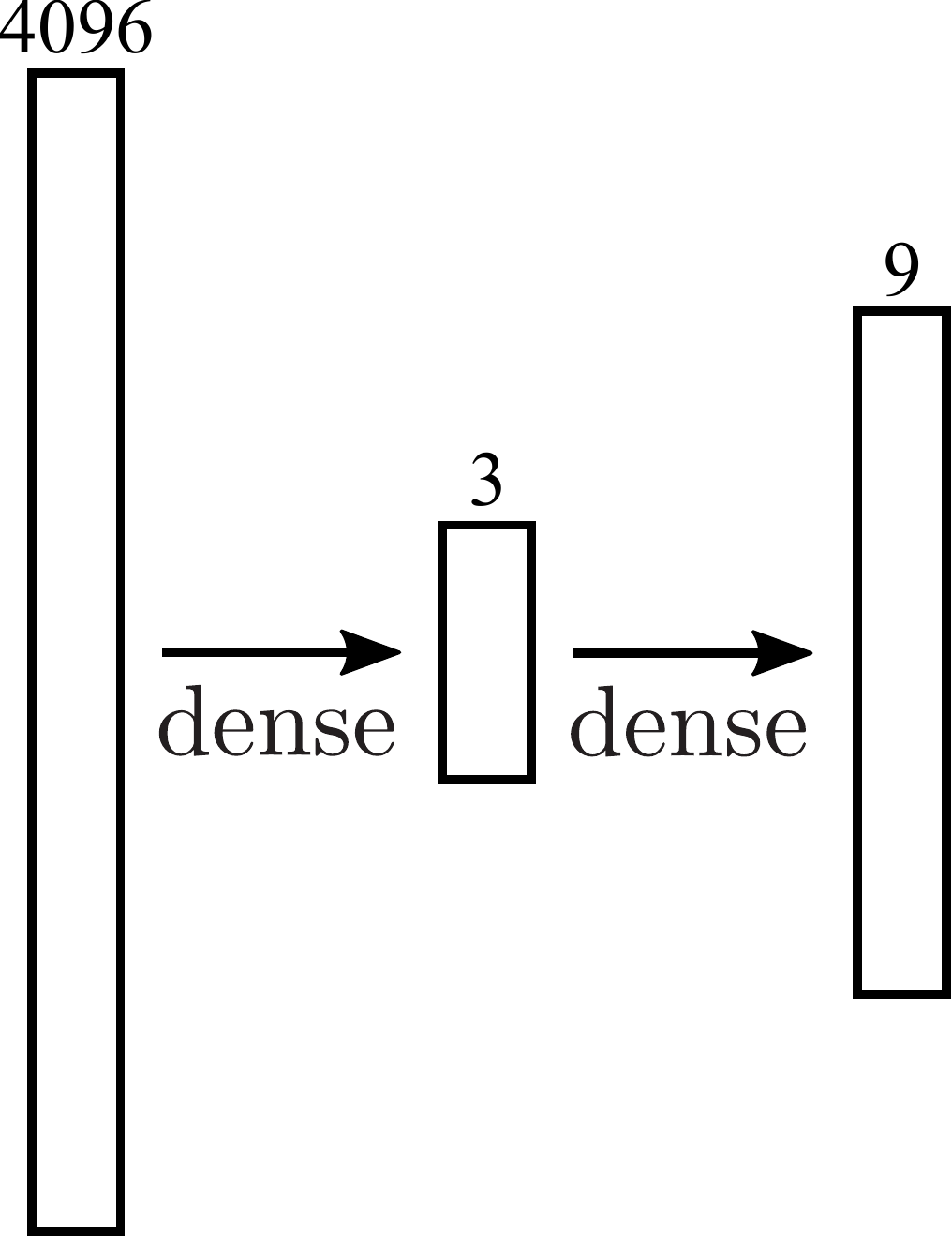}}%
  \hspace{0.1\columnwidth}
  \subfigure[Detailed structure of hidden dense layer]{%
    \includegraphics[clip, width=0.9\columnwidth]{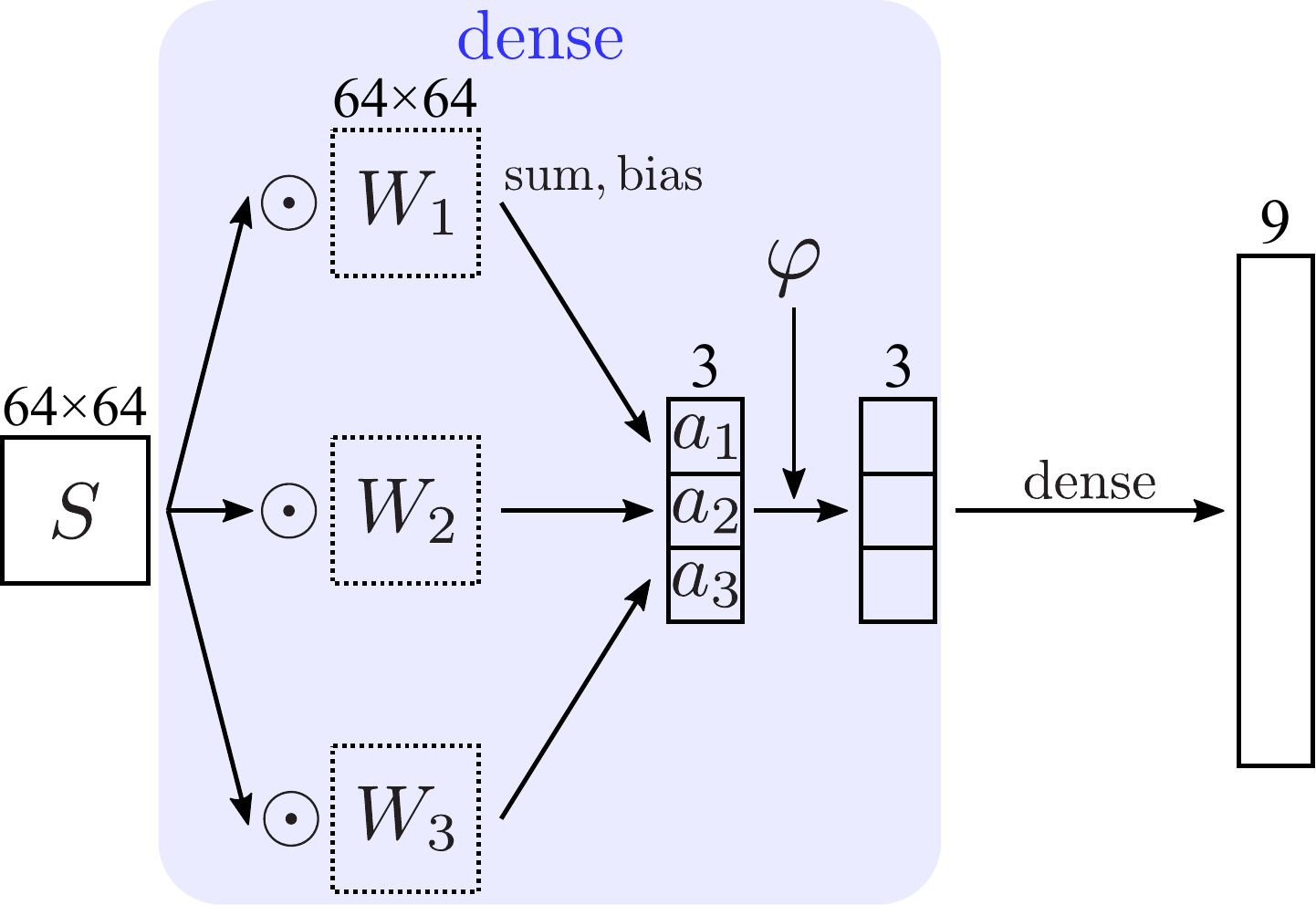}}%
  \end{center}
  \caption{\label{fig:Archi-FNN}(Color online).
    FNN architecture.
    (a) Layer structure and data sizes.
    (b) Detailed structure of the hidden dense layer, represented using $W_k$ and $a_k$ $(k=1,2,3)$.
  }
\end{figure*}

Our fully connected NN (FNN) consists of a fully connected (dense) hidden layer and an output layer (Fig.~\ref{fig:Archi-FNN}).
Both layers have the softmax activation function $\varphi : \mathbb{R}^K \to \mathbb{R}^K$~\cite{Bridle1990}:
\begin{equation}
  \varphi (\mathbf{x})_i
  = \frac{e^{x_i}}{\sum_{j=1}^{K} e^{x_j}} \quad
          (i=1,2, \ldots, K).
\end{equation}

For later discussion, we represent the weights in the hidden dense layer by three $64\times 64$ matrices $W_k$ ($k=1,2,3$) and also introduce the quantities representing the intermediate output:
\begin{equation}\label{eq:FNN-a}
  a_k=\mathrm{sum}(S\odot W_k)+\mathrm{(bias)}_k \quad (k=1,2,3)
\end{equation}
as shown in Fig.~\ref{fig:Archi-FNN}(b).
Using the intermediate output $\mathbf{a} = \{a_k\}$, the final output $\mathbf{y}$ is then given by
\begin{equation}\label{eq:FNN-y}
  \mathbf{y}=\varphi\big(
  (\mathrm{weight})\varphi(\mathbf{a})+( \mathrm{bias})
  \big).
\end{equation}
In the present paper, we do not investigate downstream of the intermediate output, for it has been discussed in detail in Ref.~\cite{Kashiwa2019}.

\subsubsection{Convolutional neural network}
\label{sec:Method-Archi-CNN}

\begin{figure*}
  \begin{center}
  \subfigure[Layer structure]{%
    \includegraphics[clip, width=0.6\columnwidth]{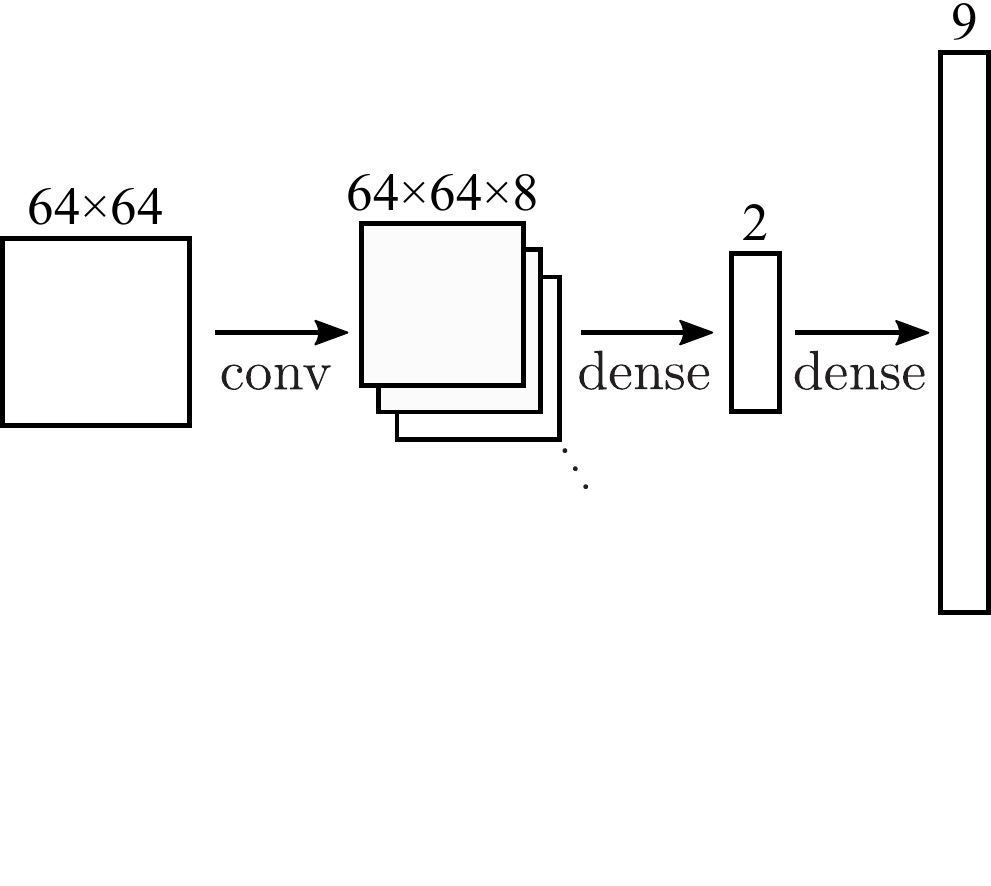}}%
  \hspace{0.1\columnwidth}
  \subfigure[Detailed structure of convolutional and hidden dense layers]{%
    \includegraphics[clip, width=1.1\columnwidth]{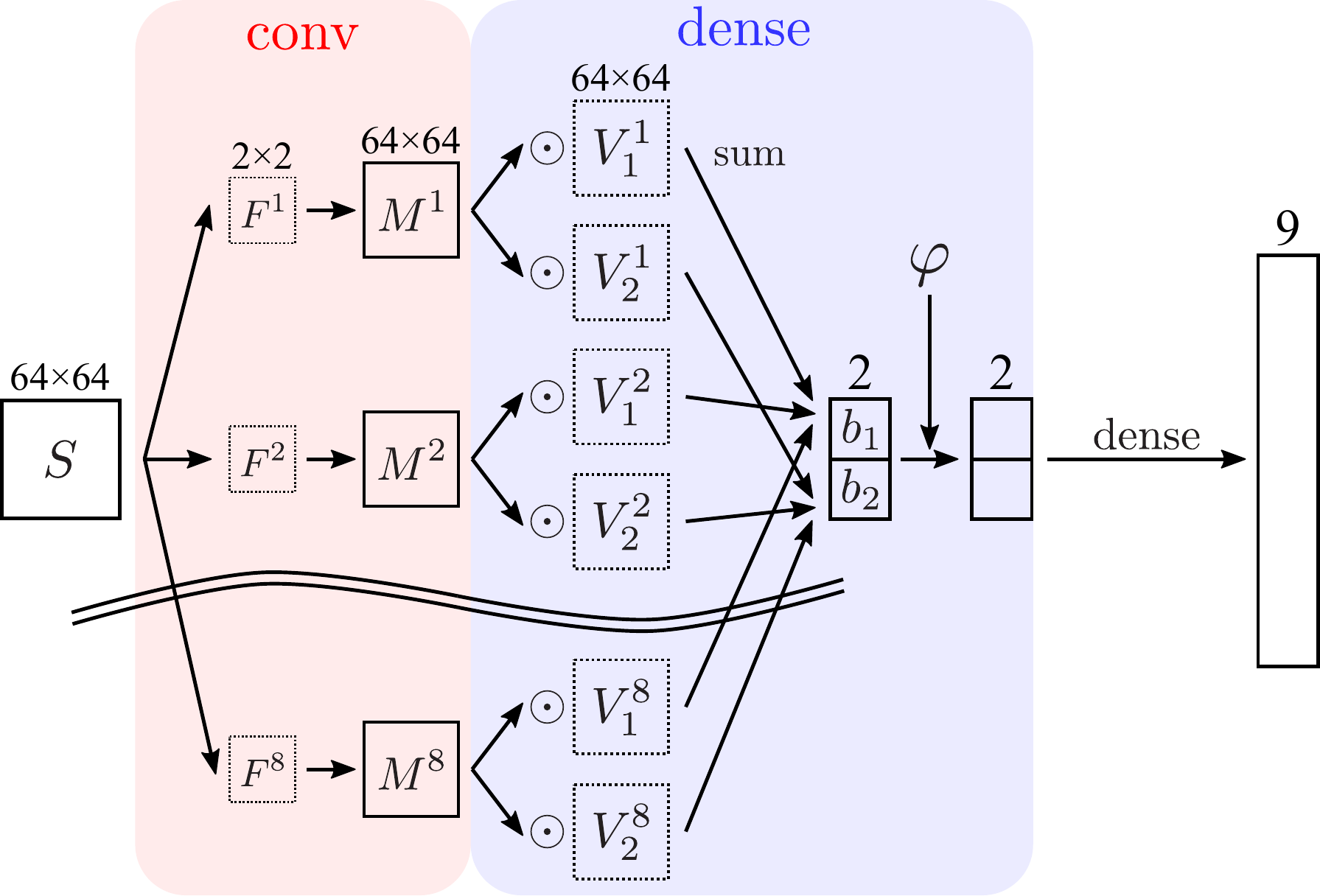}}%
  \end{center}
  \caption{\label{fig:Archi-CNN}(Color online).
    CNN architecture.
    (a) Layer structure and data sizes.
    (b) Detailed structure of the convolutional and hidden dense layers, represented using $F^f$, $M^f$, $V^f_k$ and $b_k$ ($f=1,2,\ldots,8$, $k=1,2$).
  }
\end{figure*}

Our convolutional NN (CNN) consists of the convolutional, hidden dense, and output layers as shown in Fig.~\ref{fig:Archi-CNN}.
The convolutional layer receives $S$ as an input data and generates eight feature maps $M^1, M^2,\ldots,M^{8}$. The detailed structure of convolutional layer is as follows:
\begin{itemize}
\item Eight filters $F^1, F^2,\ldots,F^{8}$ of size $2\times 2$.
\item The stride is $(1, 1)$ and the padding size is 1. The padding data is determined according to the periodic boundary conditions. By this setting, each spin is scanned exactly four times by the filters.
\item The ReLU activation function~\cite{Glorot2011,LeCun2015}:
  \begin{equation}\label{eq:CNN-ReLU}
    \mathrm{ReLU}(x)=
    \begin{cases}
      x & (x\ge 0)\\
      0 & (x<0).
    \end{cases}
  \end{equation}
\item No bias terms.
  We have confirmed that bias terms have little effect on the final results, and therefore we adopted a simpler network.
  For the same reason, we also did not use bias in the hidden dense layer.
\end{itemize}
We express $F^f$ and $M^f$ ($f=1,2,\ldots,8$) as $2 \times 2$ and $64 \times 64$
matrices, respectively.
The elements of feature maps are then given by
\begin{align}\label{eq:CNN-Map}
  \begin{split}
  M^f_{i,j}&=\mathrm{ReLU} ( \mathrm{sum}(\tilde{S}_{(i,j)}\odot F^f) )\\
  &(
  f=1,2,\ldots,8, \quad
  i,j=1,2,\ldots,64
  )
  \end{split}
\end{align}
as shown in Fig.~\ref{fig:Archi-CNN}(b).

The weights in the hidden dense layer are also represented by $64\times 64$ matrices $V^f_k$ ($k=1,2$). The intermediate output $\mathbf{b} = \{ b_k \}$ and the final output $\mathbf{y}$ are then given by
\begin{align}
  b_k&=\sum_{f=1}^{8}\mathrm{sum}(M^f\odot V^f_k) \quad (k=1,2),
  \label{eq:FNN-b} \\
  \mathbf{y}&=\varphi\big(
  (\mathrm{weight})\varphi(\mathbf{b})+( \mathrm{bias})
  \big),\label{eq:CNN-y}
\end{align}
respectively.
As in the FNN case, we do not investigate downstream of the intermediate output $\{b_k\}$.

Notice that the architecture of CNN in the present paper is different from that in the previous study~\cite{Kashiwa2019}.
We have chosen our CNN structure as simple as possible so long as there is no significant degradation in the classification accuracy for the ferromagnetic and random-gauge cases.

\section{Training Results of Neural Network}
\label{sec:Results}

In this section, we examine the intermediate output of NN and discuss the relation between the NN weights to understand how CNN extracts the physical quantities from the spin configurations of the ferromagnetic and the random-gauge models.

\subsection{Ferromagnetic Ising model}
\label{sec:Results-Ferro}

\subsubsection{Fully connected neural network}
\label{sec:Results-Ferro-FNN}

\begin{figure}[tbp]
  \begin{center}
    \includegraphics[width=0.8\hsize]{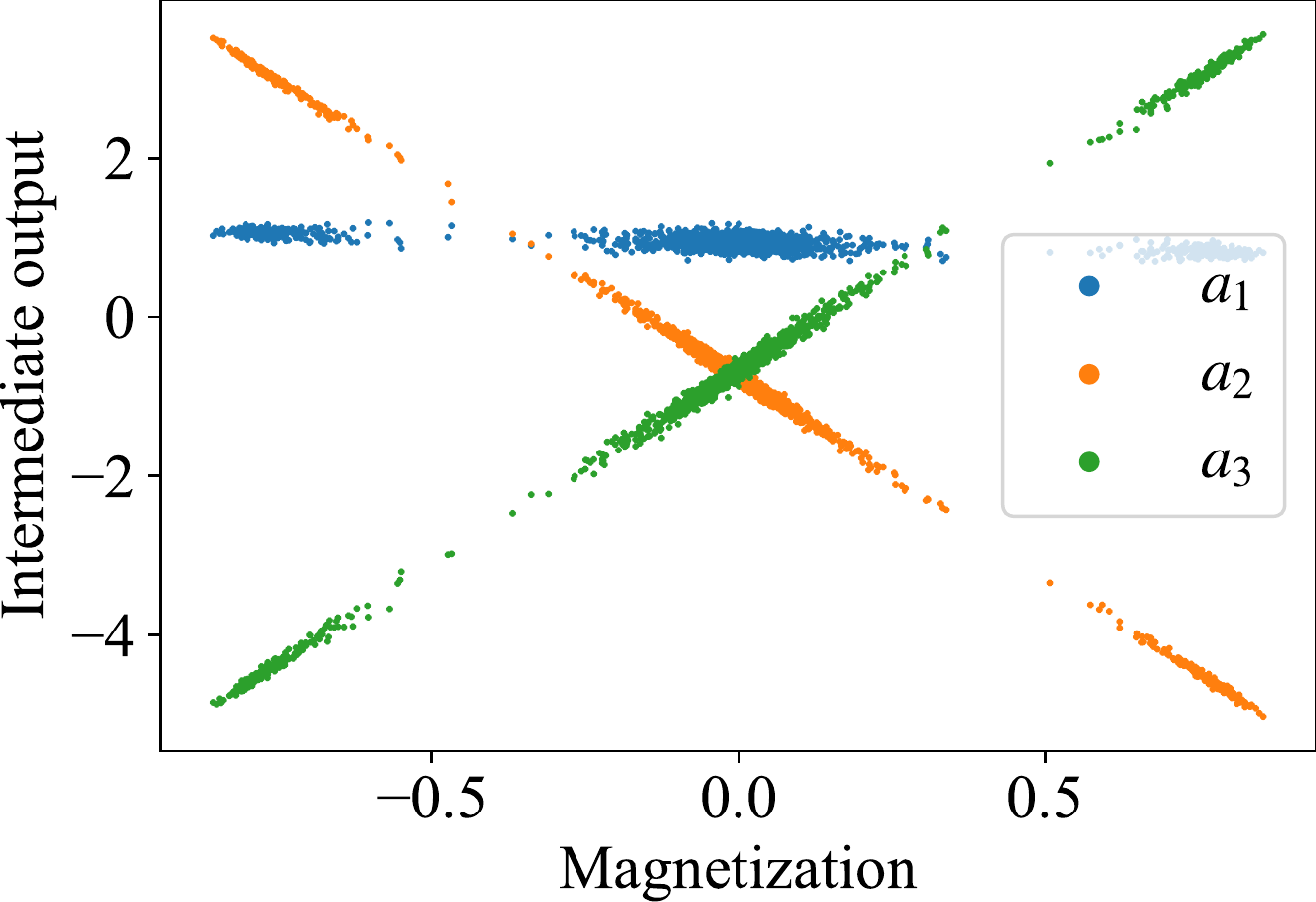}
  \end{center}
  \caption{(Color online).
    Correlation between the intermediate output $\{ a_k \}$ and the magnetization in FNN trained by the ferromagnetic dataset.
    $a_1$ is almost independent of the magnetization, while $a_2$ and $a_3$ show clear negative and positive correlations, respectively.
  }
  \label{fig:Ferro-FNN-Out}
\end{figure}

After training with the ferromagnetic dataset, FNN can infer the temperature class with a mean accuracy of 21.3\%.
In Ref.~\cite{Kashiwa2019}, it is pointed out that there is a clear correlation between the intermediate output $\{a_k\}$ and the magnetization.
Their correlation measured using the test dataset is shown in Fig.~\ref{fig:Ferro-FNN-Out}.
There are linear relations, being consistent with the previous study.
We thus conclude that the hidden dense layer of FNN successfully extracts the magnetization of the system.

\begin{figure}[tbp]
  \begin{center}
    \includegraphics[width=0.8\hsize]{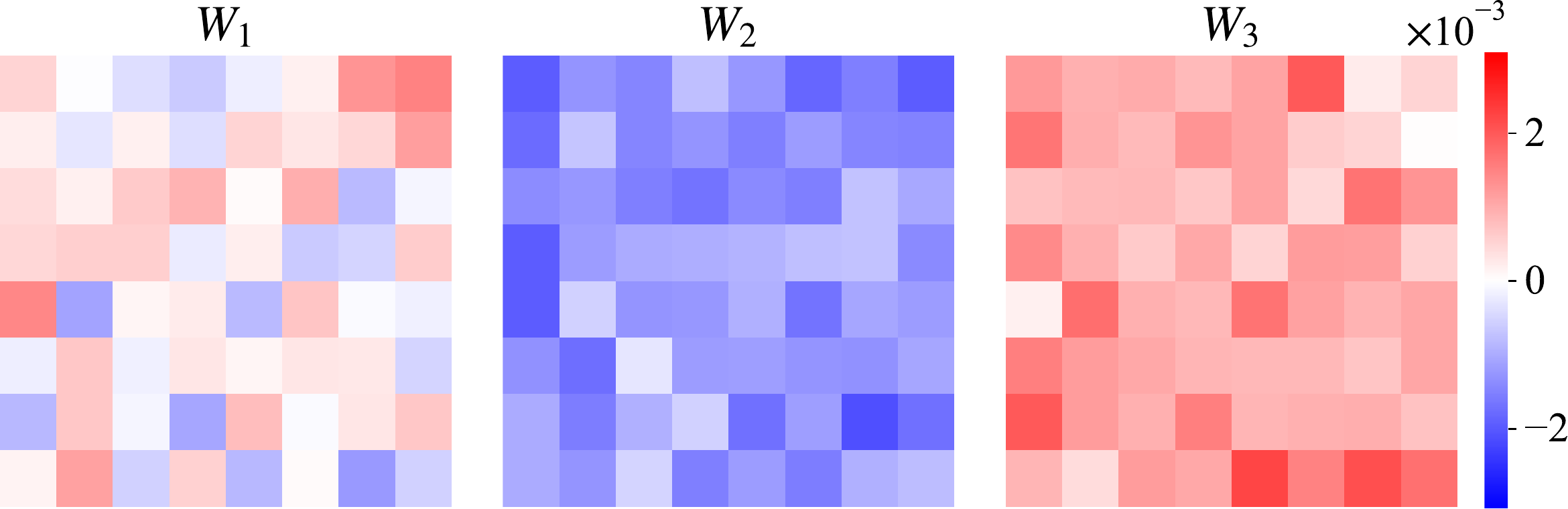}%
  \end{center}
  \caption{(Color online).
    Hidden dense weights $W_k$ ($k=1,2,3$) of FNN trained by the ferromagnetic dataset.
    Only a part ($8 \times 8$) of the whole lattice ($64 \times 64$) is presented.
    The mean of the elements of $W_1$ is close to zero, while all the elements of $W_2$ ($W_3$) have the same sign.
  }
  \label{fig:Ferro-FNN-W}
\end{figure}

To understand how the magnetization is extracted, we focus on the weights in the dense hidden layer $\{W_k\}$ (Fig.~\ref{fig:Ferro-FNN-W}).
All elements of $W_2$ (or $W_3$) have the same sign: the sum of all spin variables is calculated through these weights.
This is nothing but the (uniform) magnetization of the system.
On the other hand, the mean of the elements of $W_1$ is close to zero:
$W_1$ acts as a constant (with the bias term) to give a threshold for the critical temperature classification~\cite{Kashiwa2019}.

\subsubsection{Convolutional Neural Network}
\label{sec:Results-Ferro-CNN}

\begin{figure}[tbp]
  \begin{center}
    \includegraphics[width=0.8\hsize]{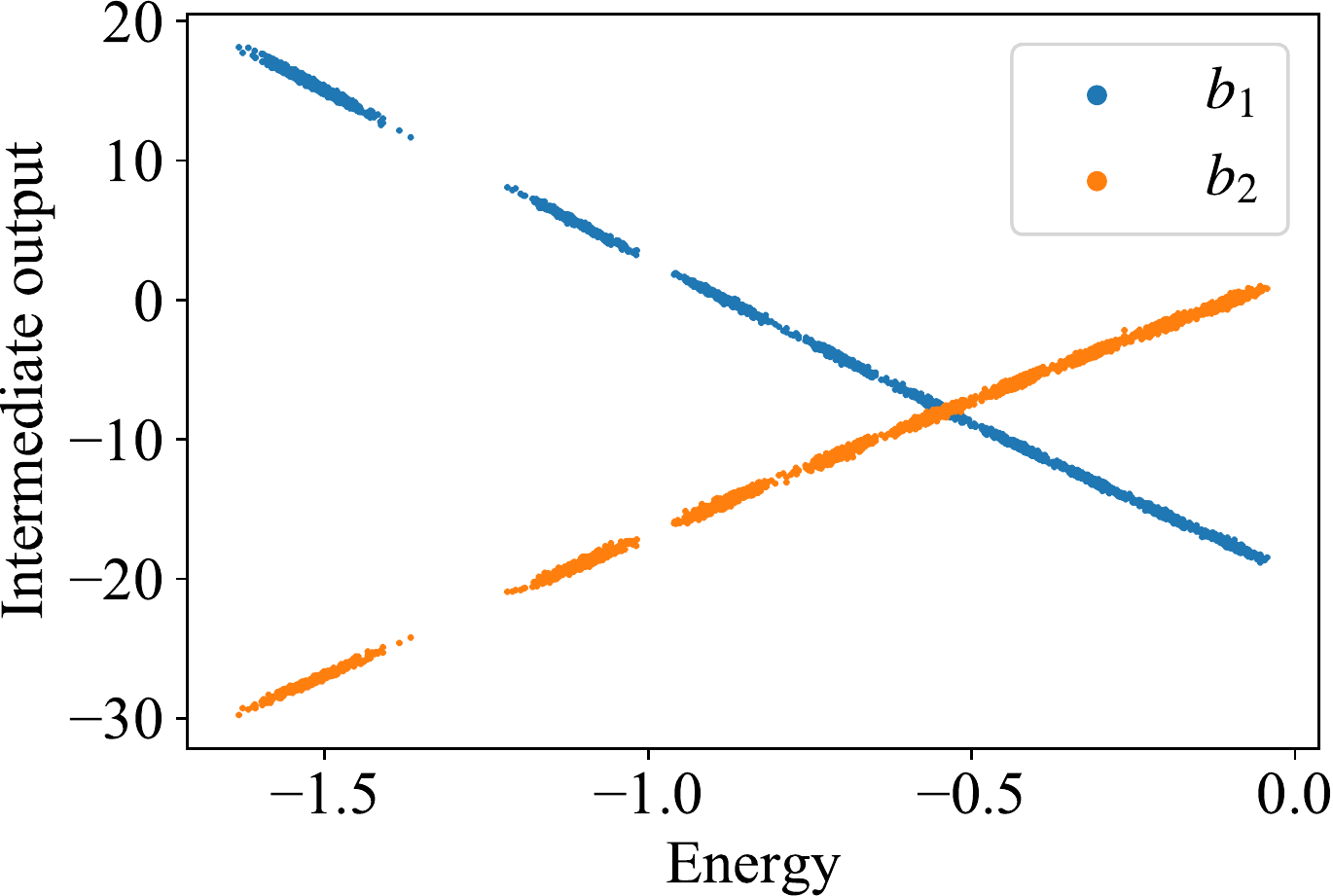}%
  \end{center}
  \caption{(Color online).
    Correlation between the intermediate output $\{b_k\}$ and the energy in CNN trained by the ferromagnetic dataset.
    $b_1$ ($b_2$) shows a clear negative (positive) correlation.
  }
  \label{fig:Ferro-CNN-Out}
\end{figure}

CNN can infer the temperature class with a mean accuracy of 95.7\% for the ferromagnetic dataset, which is significantly higher than FNN.
The correlation between the intermediate output $\{b_k\}$ and the energy for the test dataset is shown in Fig.~\ref{fig:Ferro-CNN-Out}.
We observed a clear linear relationship between $b_k$ and the energy~\cite{Kashiwa2019}, but the proportional constants strongly depend on samples and can sometimes almost vanish.
To extract the relationship between the energy and the weights in CNN more stably, we define the following quantity:
\begin{equation}\label{eq:V}
  V^f\coloneqq B_1 V^f_1+B_2V^f_2 \quad \text{($f=1,2,\ldots,8$)}.
\end{equation}
The coefficient $B_k$ ($k=1,2$) is the proportional constant between $b_k$ and the energy obtained by the least-squares fitting:
\begin{equation}\label{eq:B}
  B_k\coloneqq
  \frac{N\displaystyle\sum_i b_{i, k}E_i
    -\Big(\displaystyle\sum_i b_{i, k}\Big)
    \Big(\displaystyle\sum_i E_i\Big)}
       {N\displaystyle\sum_i E_i^2-\Big(\displaystyle\sum_i E_i\Big)^2} \quad \text{($k=1,2$)},
\end{equation}
where $i$ represents the sample index, $E_i$ and $b_i$ are its energy and intermediate output respectively, and $N$ is the total number of test data.

\begin{figure}[tbp]
  \begin{center}
    \includegraphics[width=\hsize]{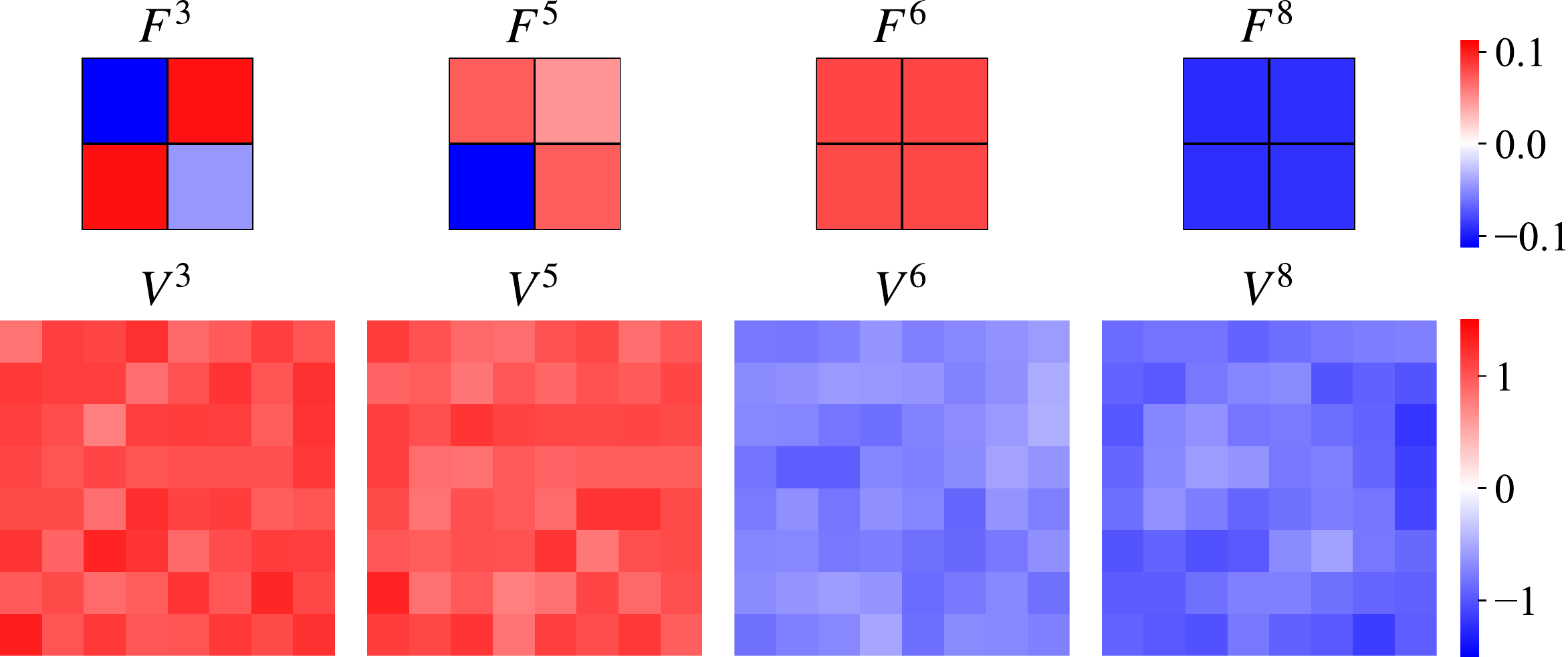}%
  \end{center}
  \caption{(Color online).
    Selected convolutional filters $F^f$ and corresponding weights $V^f$ of CNN trained by the ferromagnetic dataset.
    For $V^f$, only a part ($8 \times 8$) of the whole lattice ($64 \times 64$) is presented.
    If all the elements of the filters have the same sign ($F^6$ and $F^8$), the corresponding weights ($V^6$ and $V^8$) have elements with negative sign.
    Otherwise ($F^3$ and $F^5$), the weights ($V^3$ and $V^5$) have elements with positive sign.
  }
  \label{fig:Ferro-CNN-FV}
\end{figure}

Selected convolutional filters $F^f$ and their corresponding weights $V^f$ are shown in Fig.~\ref{fig:Ferro-CNN-FV}, where we can see the following features:
\begin{enumerate}[label=(\alph*)]
  \item In each $V^f$, all the elements have the same sign.
  \item If all the elements of $F^f$ have the same sign, the elements of $V^f$ are negative.
    Otherwise, they are positive.
\end{enumerate}
The weights $\{V^f\}$ are thus related to the local energy of the spin configurations represented by the corresponding convolutional filters $\{F^f\}$.
To be more specific, if the energy of the spin configuration $F^f$ is -4 (e.g., $f=6,8$ in Fig.~\ref{fig:Ferro-CNN-FV}), the elements of $V^f$ are negative; on the other hand, if the energy of $F^f$ is +4 ($f=3$) or 0 ($f=5$), $V^f$ has positive elements.
Notice that the filter pattern and the values of $V^f$ do not have one-to-one correspondence.
This is because a filter detects the same spin patterns as itself and the ones slightly different from itself, and such patterns possibly have multiple energy values.
We conclude that CNN extracts the energy by detecting the local spin configuration by $F^f$ and assigning $V^f$ as its contribution to the total energy.

\subsection{Random-gauge Ising model}
\label{sec:Results-RG}

\subsubsection{Fully connected neural network}
\label{sec:Results-RG-FNN}

\begin{figure}[tbp]
  \begin{center}
    \includegraphics[width=0.8\hsize]{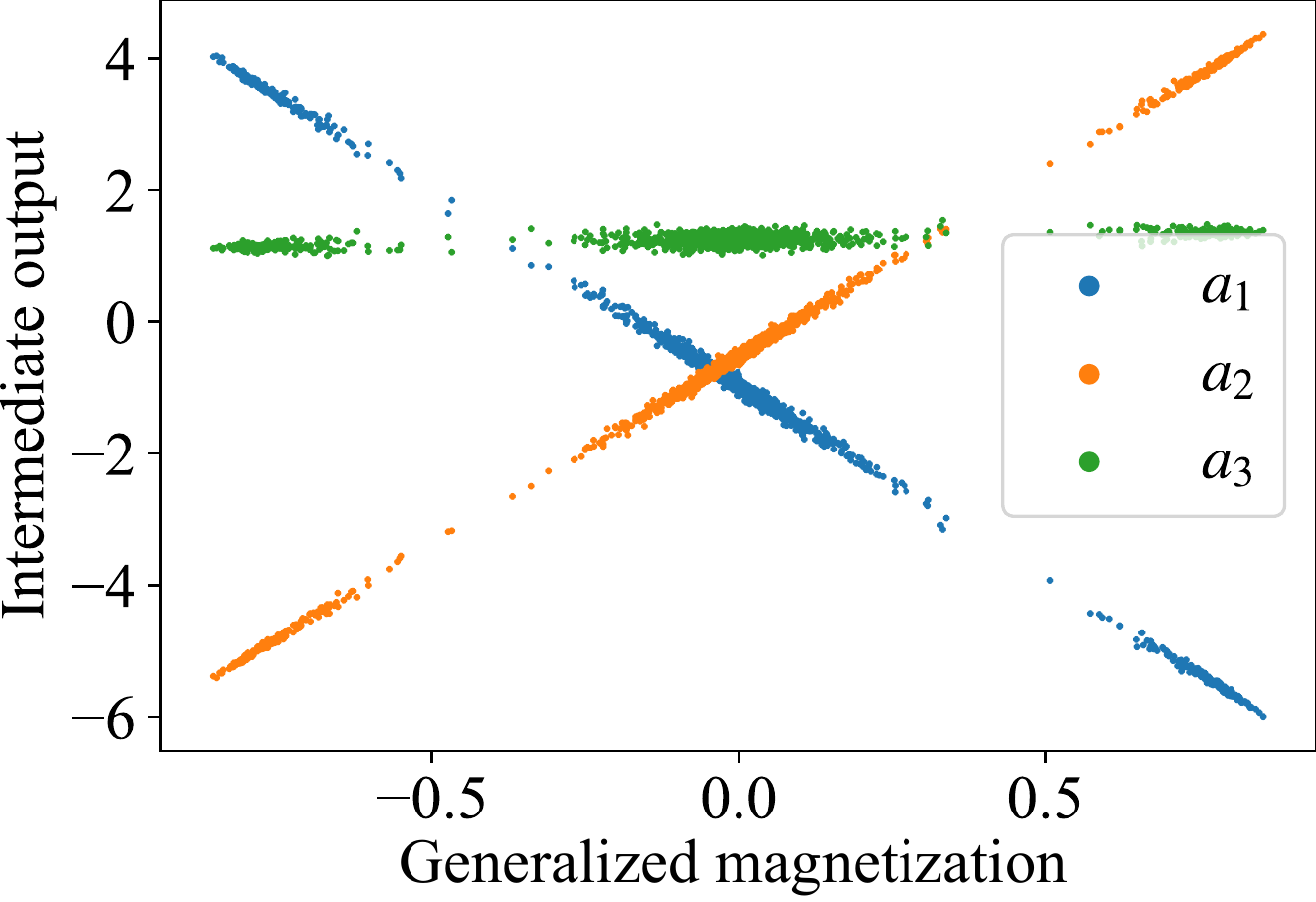}%
  \end{center}
  \caption{(Color online).
    Correlation between the intermediate output $\{a_k\}$ and the generalized magnetization in FNN trained by the random-gauge dataset.
    $a_3$ is almost independent of the generalized magnetization, while $a_1$ and $a_2$ show clear negative and positive correlations, respectively.
  }
  \label{fig:RG-FNN-Out}
\end{figure}

\begin{figure}[tbp]
  \begin{center}
    \includegraphics[width=\hsize]{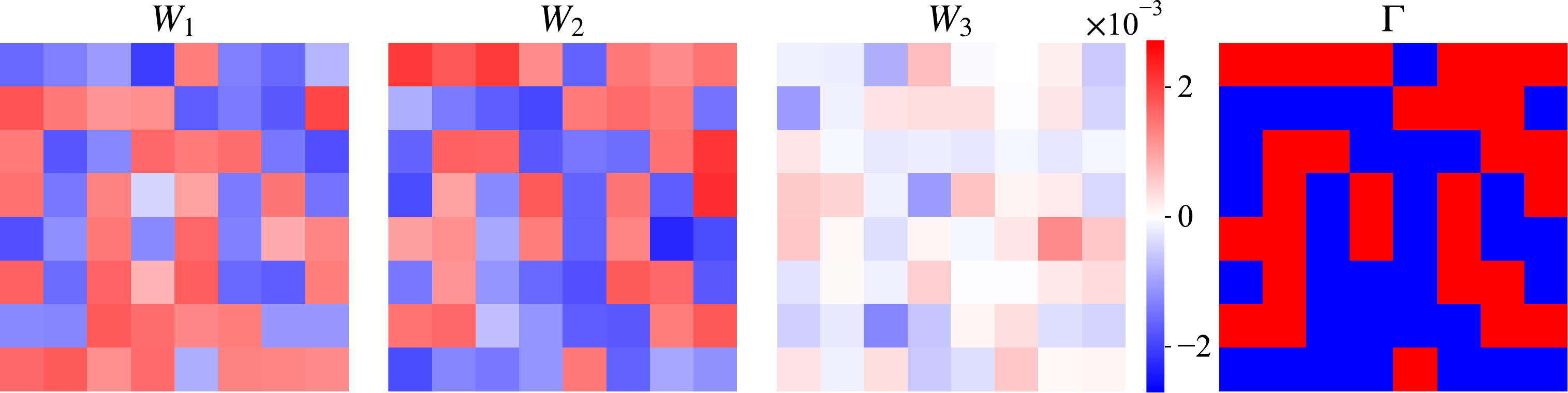}%
  \end{center}
  \caption{(Color online).
    Hidden dense weights $W_k$ ($k=1,2,3$) of FNN trained by the random-gauge dataset.
    The gauge $\Gamma $ is also shown for comparison.
    Only a part ($8 \times 8$) of the whole lattice ($64 \times 64$) is presented.
    The mean of the elements of $W_3$ is close to zero, while $W_2$ and $W_3$ have the same pattern as $\Gamma$.
  }
  \label{fig:RG-FNN-W}
\end{figure}

FNN has a mean accuracy of 21.5\% for the random-gauge dataset, almost the same as the ferromagnetic case.
The correlation between the intermediate output $\{a_k\}$ and the magnetization is shown in Fig.~\ref{fig:RG-FNN-Out}.
Notice that the horizontal axis represents the generalized magnetization~\eqref{eq:generalized-magnetization}.
The behavior of the intermediate output $\{a_k\}$ is similar to the ferromagnetic case (Fig.~\ref{fig:Ferro-FNN-Out}(b)).
It indicates that FNN can extract the generalized magnetization correctly without being confused by the gauge.
To find out how it is done, we take a close look at the hidden dense weights $\{W_k\}$ (Fig.~\ref{fig:RG-FNN-W}).
The weight $W_3$ has the elements close to zero, just like $W_1$ in the ferromagnetic case (Fig.~\ref{fig:Ferro-FNN-W}).
In contrast, the elements of $W_1$ and $W_2$, unlike the ferromagnetic the case, do not have the same sign.
Careful observation reveals that $W_1$ and $W_2$ have the same pattern as $\Gamma $, implying that the hidden dense layer learns the gauge $\Gamma $ from the training dataset and extracts the generalized magnetization successfully as an order parameter of the random-gauge model.

\subsubsection{Convolutional neural network}
\label{sec:Results-RG-CNN}

\begin{figure}[tbp]
  \begin{center}
    \includegraphics[width=0.8\hsize]{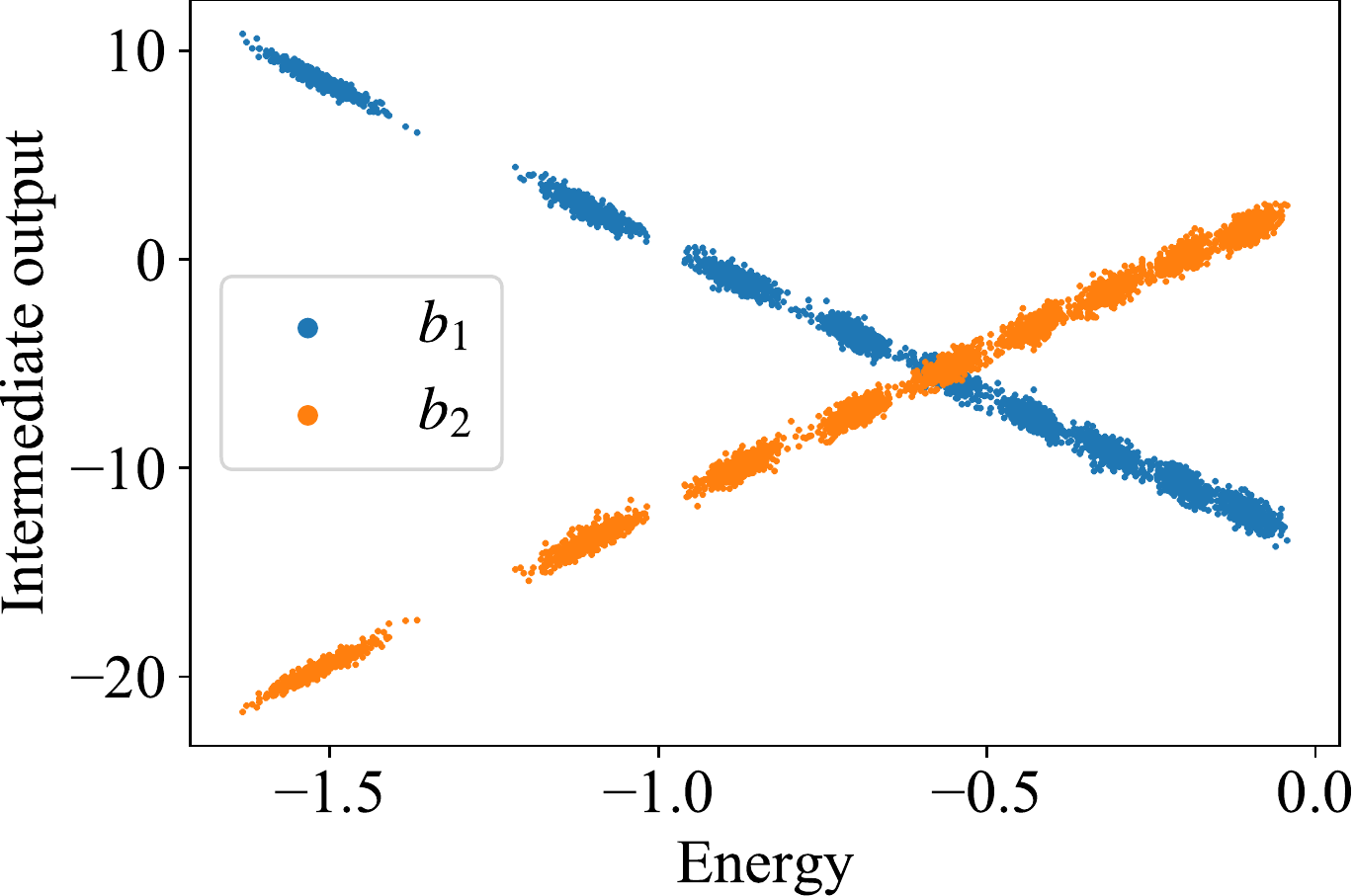}%
  \end{center}
  \caption{(Color online).
    Correlation between the intermediate output $\{b_k\}$ and the energy in CNN trained by the random-gauge dataset.
    $b_1$ ($b_2$) shows a clear negative (positive) correlation.
  }
  \label{fig:RG-CNN-Out}  
\end{figure}

\begin{figure}[tbp]
  \begin{center}
    \includegraphics[width=\hsize]{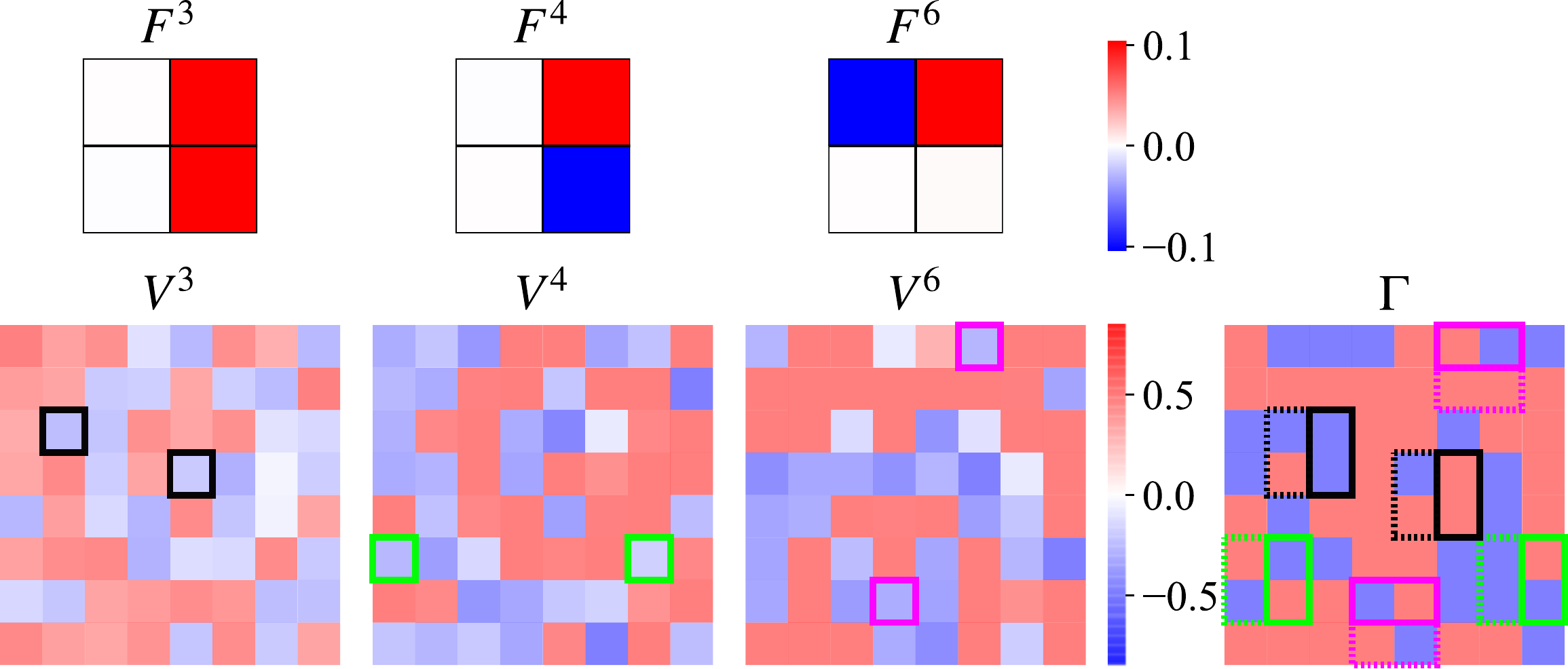}%
  \end{center}
  \caption{(Color online).
    Selected convolutional filters $F^f$ and corresponding hidden dense weights $V^f$ of CNN trained by the random-gauge dataset.
    The gauge $\Gamma$ is also shown for comparison.
    For $V^f$ and $\Gamma$, only a part ($8 \times 8$) of the whole lattice ($64 \times 64$) is presented.
    Some corresponding elements of $V^f$ and $\Gamma $ are framed.
    Solid/dashed line corresponds to non-zero/zero elements of the filters.
  }
  \label{fig:RG-CNN-FV}
\end{figure}

CNN has a mean accuracy of 94.2\% for the random-gauge dataset, slightly lower than the ferromagnetic case.
The correlation between the intermediate output $\{b_k\}$ and the energy graph is shown in Fig.~\ref{fig:Ferro-CNN-Out}.
The intermediate output $\{b_k\}$ behaves like that in the ferromagnetic case (Fig.~\ref{fig:Ferro-FNN-Out}(b)), which indicates that CNN can extract the energy of the random-gauge model without being confused by the gauge.

Selected convolutional filters $\{F^f\}$ and their corresponding weights $\{V^f\}$ are shown in Fig.~\ref{fig:RG-CNN-FV}.
We observed the following features:
\begin{enumerate}[label=(\alph*)]
   \item \label{enum:a}
      In each $V^f$, elements do not have the same sign, but they are split into two classes, positive and negative.
   \item \label{enum:b}
      In each $F^f$, only two are finite among four elements.
      The absolute value of the other two elements is much smaller.
      We regard the latter two as \textit{zero}.
   \item \label{enum:c}
      If non-zero elements of $F^f$ have the same pattern as the local gauge ($\tilde{\Gamma }_{(i,j)}$), the element of $V^f$ on the corresponding site ($V^f_{i,j}$) is negative, otherwise positive (see the framed boxes in Fig.~\ref{fig:RG-CNN-FV}).
\end{enumerate}
We have confirmed that all the other filters and weights have the same feathers.
Careful consideration about the feature~\ref{enum:c} reveals that CNN extracts the local energy in the same way as the ferromagnetic case: the compatibility (incompatibility) of local spins and gauge leads to low (high) energy.
The feature~\ref{enum:a} indicates CNN learns that the same spin pattern gives different energy depending on the position in the lattice in the random-gauge case.
Interestingly, the feature~\ref{enum:b} implies CNN learns that for calculating the energy, it is more efficient to look at the relationship between two spins than looking at four spins simultaneously in the random-gauge case.
CNN has to prepare at least eight different filters if four elements are non-zero depending on the gauge patterns.
On the other hand, four (two for horizontal and two for vertical pairs) filters are enough if each filter has only two non-zero elements.

\section{Gauge Reconstruction}
\label{sec:Gauge}

In Sec.~\ref{sec:Results}, we have discussed how the information of the gauge is encoded in the NN weights.
In this section, we attempt to reconstruct the gauge from the NN parameters to demonstrate the validity of the discussion in the previous section.
Notice that since we did not include low-temperature data in the training dataset (see Sec.~\ref{sec:Method-Dataset}), no or few configurations are identical to the gauge itself.
We denote the reconstructed gauge as $\Gamma_\mathrm{rec}$ or $\Theta _ \mathrm{rec}$ in the following.

\subsection{Fully connected neural network}
\label{sec:Gauge-FNN}

Based on the observation in Sec.~\ref{sec:Results-RG}, we expect that the gauge can be reconstructed from the difference between $W_1$ and $W_2$ as
\begin{equation}\label{eq:Gauge-FNN}
  \Gamma _ \mathrm{rec}= \pm \mathrm{sgn}(W_1-W_2).
\end{equation}
We define the reconstruction error by the Frobenius norm of $(\Gamma _ \mathrm{rec} - \Gamma)/2$, i.e., the number of sites with incorrect gauge.
Note that we can not determine the overall sign in Eq.~(\ref{eq:Gauge-FNN}) from the weights due to the $\mathbb{Z}_2$ symmetry of the Ising model.
We chose the sign so that the error becomes smaller.

\begin{figure}[tbp]
  \begin{center}
    \includegraphics[width=0.8\hsize]{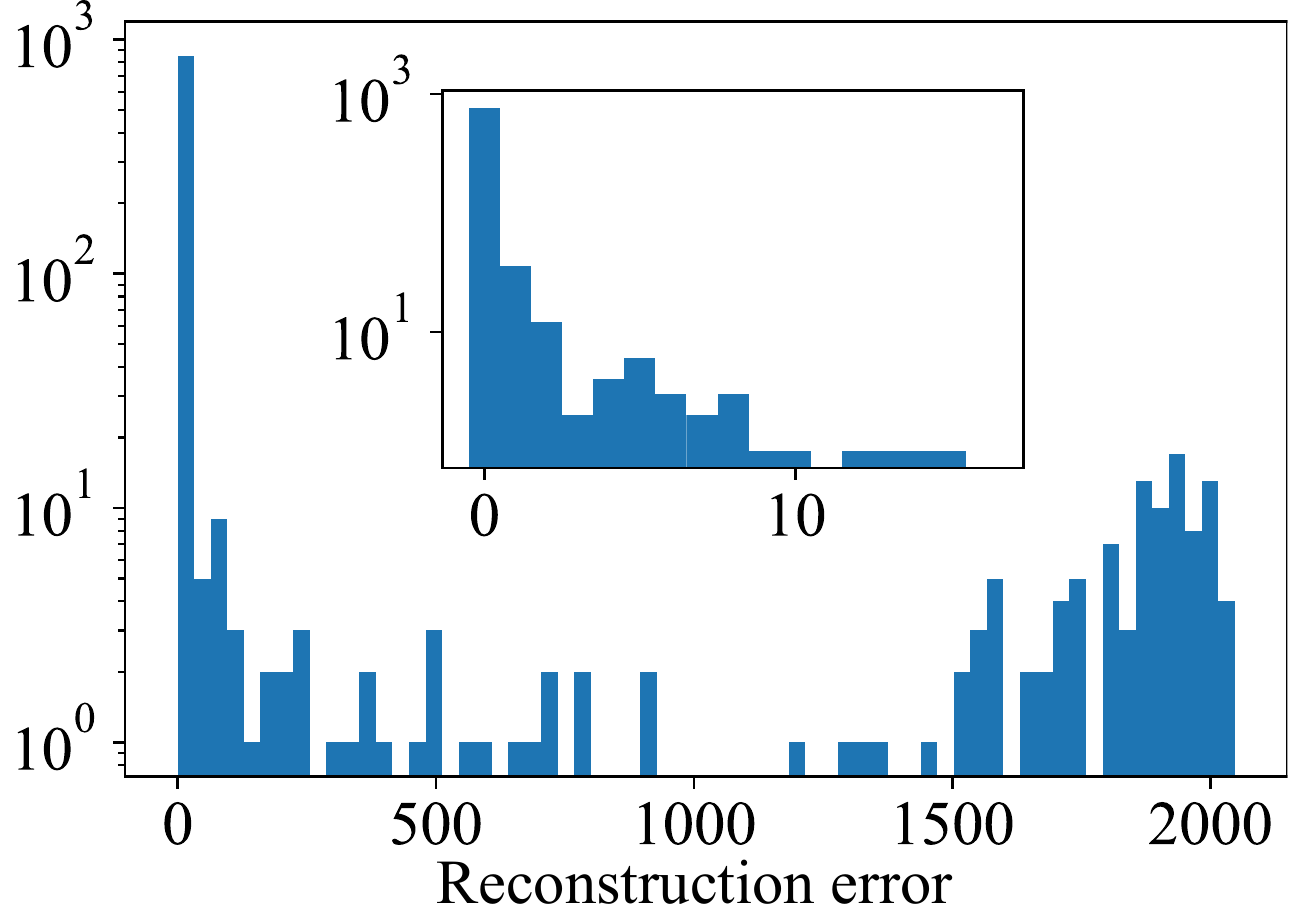}%
  \end{center}
  \caption{(Color online).
    Histogram of gauge-reconstruction error for FNN.
    The total number of trials is 1000.
    The vertical axis is plotted on a logarithmic scale as the histogram has a sharp peak at zero.
    The upper bound of the error is $L^2/2 = 2048$ by definition.
    The bin size is 32 in the main plot.
    The inset shows the histogram near zero with bin size 1.
  }
  \label{fig:FNN-Hist}
\end{figure}

We carried out the gauge reconstruction for 1000 different gauge patterns.
The histogram of the reconstruction error of 1000 trials is shown in Fig.~\ref{fig:FNN-Hist}.
We found that more than 75\% of trials achieve perfect gauge reconstruction with no errors, supporting that the hidden dense weights $\{W_k\}$ of FNN encode the gauge information precisely.

In Fig.~\ref{fig:FNN-Hist}, we should notice that a certain number of trials yield the error around 2000.
We have confirmed that the $a_k$-magnetization graph does not show a linear relationship in such cases, indicating that FNN is trapped at a local minimum and fails to extract the generalized magnetization.
As a result, the reconstructed gauge gives a random pattern that does not correlate with the real one.
Even in such cases learning finishes successfully with different initial conditions, and so does the gauge reconstruction.

\subsection{Convolutional neural network}
\label{sec:Gauge-CNN}

As we have seen in Sec.~\ref{sec:Results-RG}, $F^f$ and $\tilde{\Gamma }_{(i,j)}$ have the same pattern if $V^f_{i,j}$ is negative, and vice versa.
One may expect that we can reconstruct the gauge by embedding the same (or different) pattern as $F^f$ into $(\tilde{\Gamma }_\mathrm{rec})_{(i,j)}$ according to the sign of $V^f_{i,j}$.
However, this reconstruction strategy does not work, for both $+F^f$ and $-F^f$ correspond to the same sign of $V^f$ (e.g., $f=6,8$ in Fig.~\ref{fig:Ferro-CNN-FV}) and hence simple embedding of $F^f$ may cause canceling each other.
Instead, we convert $F^f$ into local coupling constant $J^f$ in advance and then reconstruct $\Theta$ instead of $\Gamma$.
It solves the canceling problem as both $+F^f$ and $-F^f$ yield the same $J^f$.

\begin{figure}[tbp]
  \begin{center}
    \includegraphics[width=0.8\hsize]{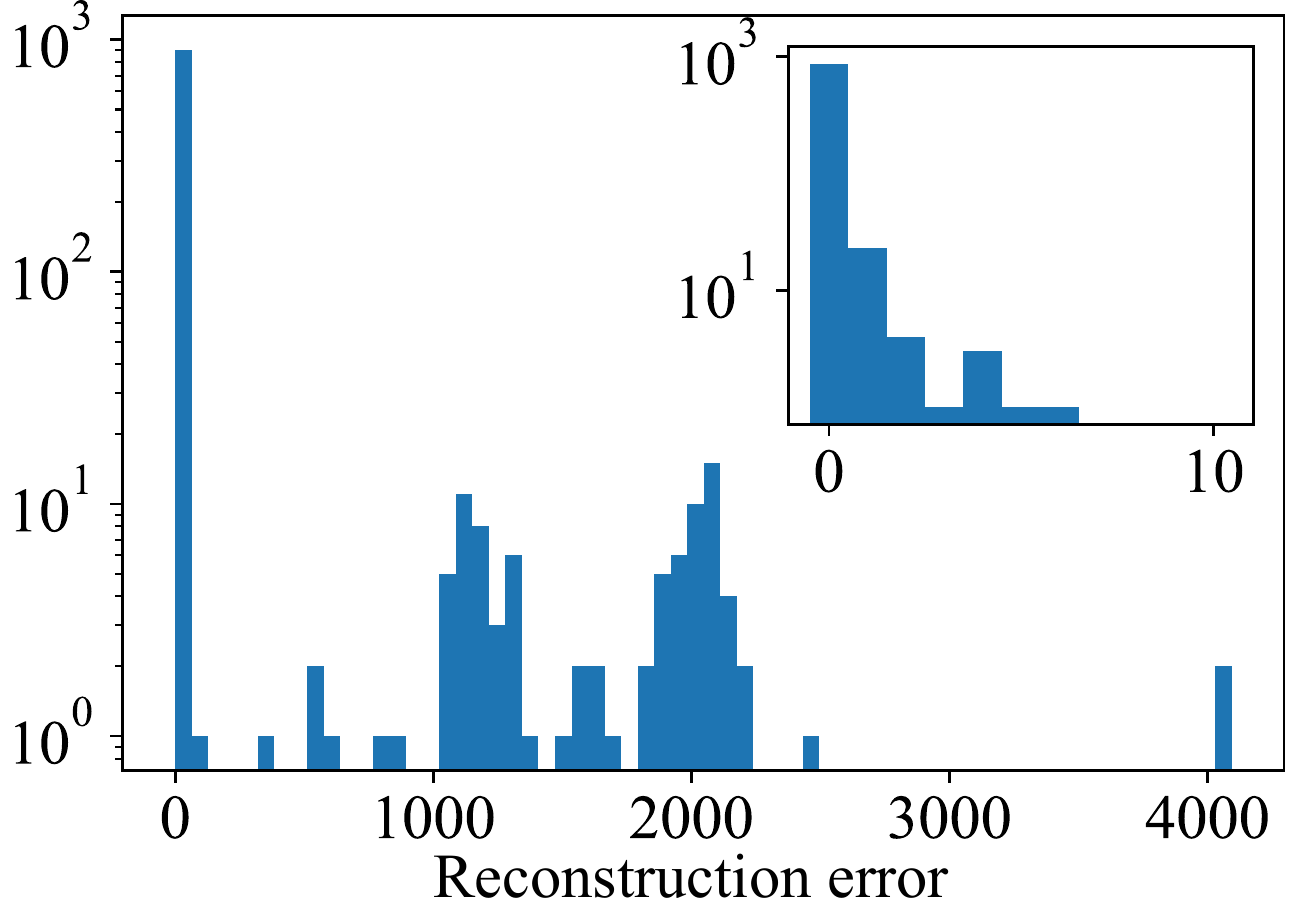}%
  \end{center}
  \caption{(Color online).
    Histogram of gauge-reconstruction error for CNN.
    The total number of trials is 1000.
    The vertical axis is plotted on a logarithmic scale as the histogram has a sharp peak at zero.
    The upper bound of the number of errors is $L^2=4096$.
    The bin size is 64 in the main plot.
    The inset shows the histogram near zero with bin size 1.
  }
  \label{fig:CNN-Hist}        
\end{figure}

The reconstruction procedure of $\Theta_\mathrm{rec}$ is as follows:
\begin{enumerate}
  \item Convert $F^f$ into local coupling constants $J^f$ by taking a product of the adjacent elements.
  \item Initialize $\Theta_\mathrm{rec}$ to zero.
  \item Add $-V^f_{i,j} J^f$ into the corresponding position of $\Theta_\mathrm{rec}$.
  \item Repeat step 3 for all $f$ and $(i,j)$.
  \item Take the sign of each element:
    \begin{equation}
      \Theta_\mathrm{rec}\leftarrow \mathrm{sgn}(\Theta_\mathrm{rec}). \notag
    \end{equation}
\end{enumerate}
As in the FNN case, we carried out the reconstruction 1000 times with different gauge patterns and initial values of NN parameters.
The result is shown in Fig.~\ref{fig:CNN-Hist}.
We found that more than 90\% trials achieve perfect gauge reconstruction, supporting that the convolutional filters $\{F^f\}$ and the hidden dense weights $\{V^f\}$ encode the local gauge patterns and their energies, respectively.

In Fig.~\ref{fig:CNN-Hist}, we should notice that a certain number of trials yield the error around 2000.
We have confirmed that in such cases, no filters have non-zero vertically (horizontally) aligned elements.
As a result, it becomes impossible to reconstruct vertical (horizontal) coupling constants, and
about a quarter of the coupling constants become an error.
Just as in the FNN case, learning finishes successfully with different initial conditions for NN parameters, even in such cases.

\section{Multiple Gauges}
\label{sec:Mult}

We have investigated the random-gauge model, in which configurations are shuffled by a single gauge so far.
One may ask how NN performs if the dataset consists of spin configurations perturbed with two or more different gauges.
This section presents the result of experiments for the simplest case, the random-gauge model with \textit{two} gauge patterns.
The experimental setup is the same as before, except we apply two different random gauges $\Gamma_1$ and $\Gamma _2$ ($\Theta_1$ and $\Theta _2$), which are statistically independent, for the first and second halves of the dataset, respectively.

\subsection{Fully connected neural network}
\label{sec:Mult-FNN}

\begin{figure}[tbp]
  \begin{center}
    \includegraphics[width=0.8\hsize]{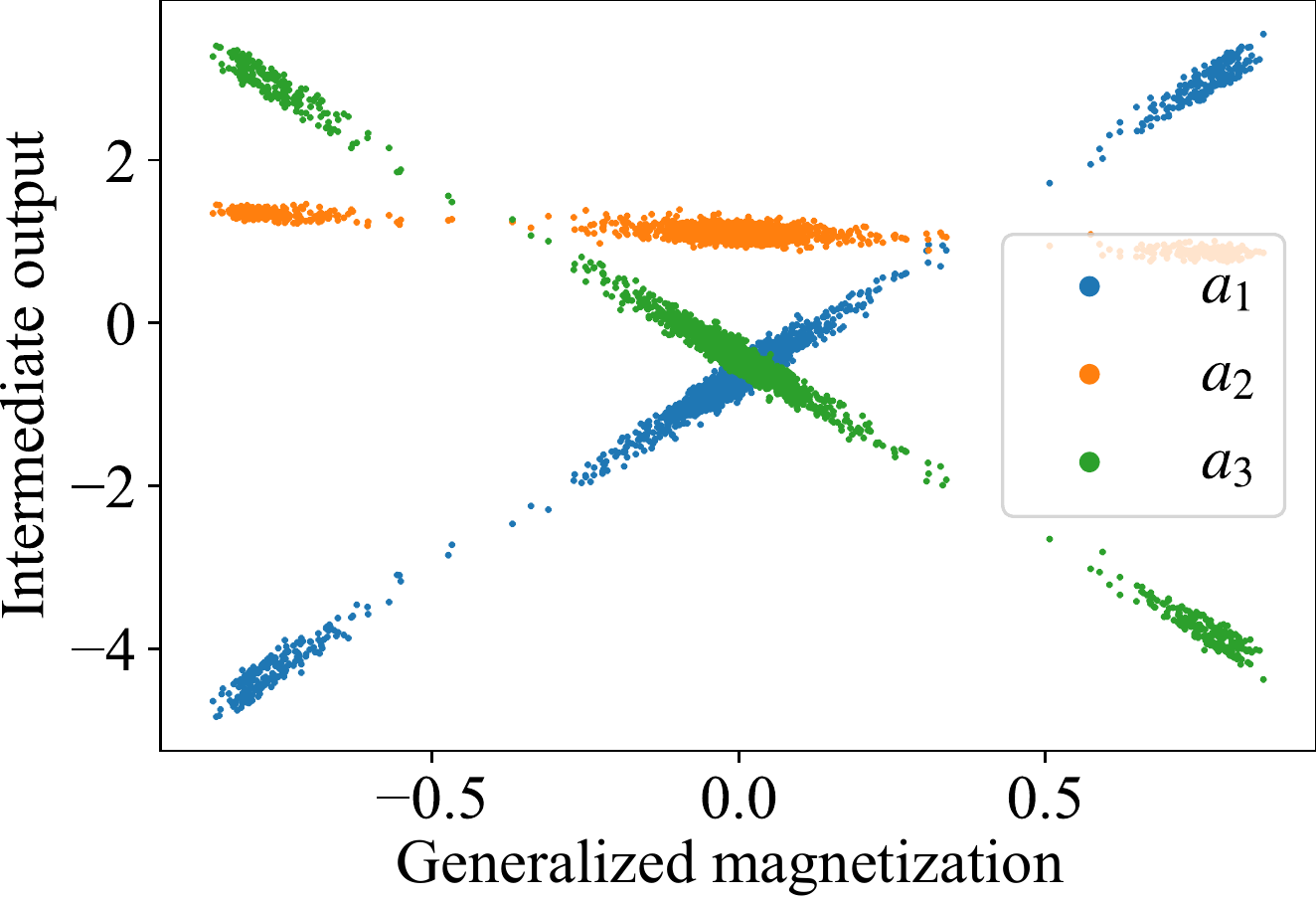}%
  \end{center}
  \caption{(Color online).
    Correlation between the intermediate output $\{ a_k \}$ and the generalized magnetization in FNN trained by the double-random-gauge dataset.
    $a_2$ is almost independent of the magnetization, while $a_1$ and $a_3$ show clear positive and negative correlations, respectively.
  }
  \label{fig:G2-FNN-Out}
\end{figure}

\begin{figure}[tbp]
  \begin{center}
    \includegraphics[width=0.7\hsize]{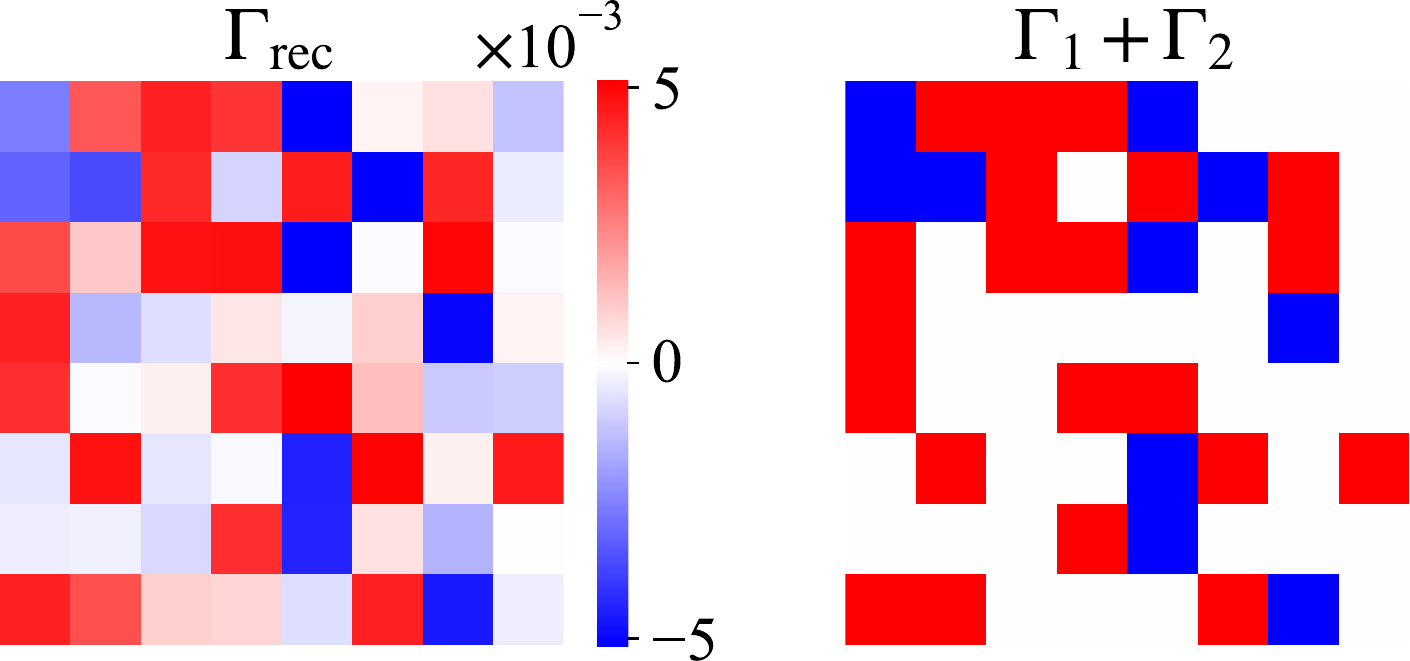}%
  \end{center}
  \caption{(Color online).
    Reconstructed gauge $\Gamma_\mathrm{rec}$ from FNN trained by the double-random-gauge dataset. For comparison, we also show the sum of two gauges $\Gamma _1+\Gamma _2$.
    Only a part ($8 \times 8$) of the whole lattice ($64 \times 64$) is presented.
  }
  \label{fig:G2-FNN-G}
\end{figure}

The mean classification accuracy is $18.6\%$, which is not that much worse than the single gauge case.
The correlation between the intermediate output $\{a_k\}$ and the generalized magnetization is shown in Fig.~\ref{fig:G2-FNN-Out}.
It can be seen that although fluctuations of $\{a_k\}$ become larger than the previous case, FNN successfully extracts the generalized magnetization.
We have tried to reconstruct the gauge $\Gamma_ \mathrm{rec}$ according to the same procedure proposed in Sec.~\ref{sec:Gauge-FNN} except that we did not apply the sign function [Eq.~(\ref{eq:Gauge-FNN})] this time.
We found that $\Gamma_\mathrm{rec}$ has a similar pattern as $\Gamma _1 + \Gamma _2$ as shown in Fig.~\ref{fig:G2-FNN-G}.

More quantitatively, the mean correlation coefficient between the elements of $\Gamma_ \mathrm{rec}$ and $\Gamma _1 + \Gamma _2$ for 1000 trials is evaluated as 0.725.
If we look at each trial closely, we find two categories: the reconstructed gauges with correlation coefficient almost unity and those with almost zero.
We have confirmed that in the latter cases, there is no significant correlation between $\{a_k\}$ and the generalized magnetization, indicating failure in learning from the dataset.
On the other hand, in the former successful cases, we conclude that FNN detects lattice sites where $\Gamma _1$ and $\Gamma _2$ have the same sign and uses them for temperature prediction.
The increase of fluctuations of $\{a_k\}$ than the single gauge case can also be understood by considering that FNN uses only about a half of spin configurations to extract the order parameter.

\subsection{Convolutional neural network}
\label{sec:Mult-CNN}

\begin{figure}[tbp]
  \begin{center}
    \includegraphics[width=0.8\hsize]{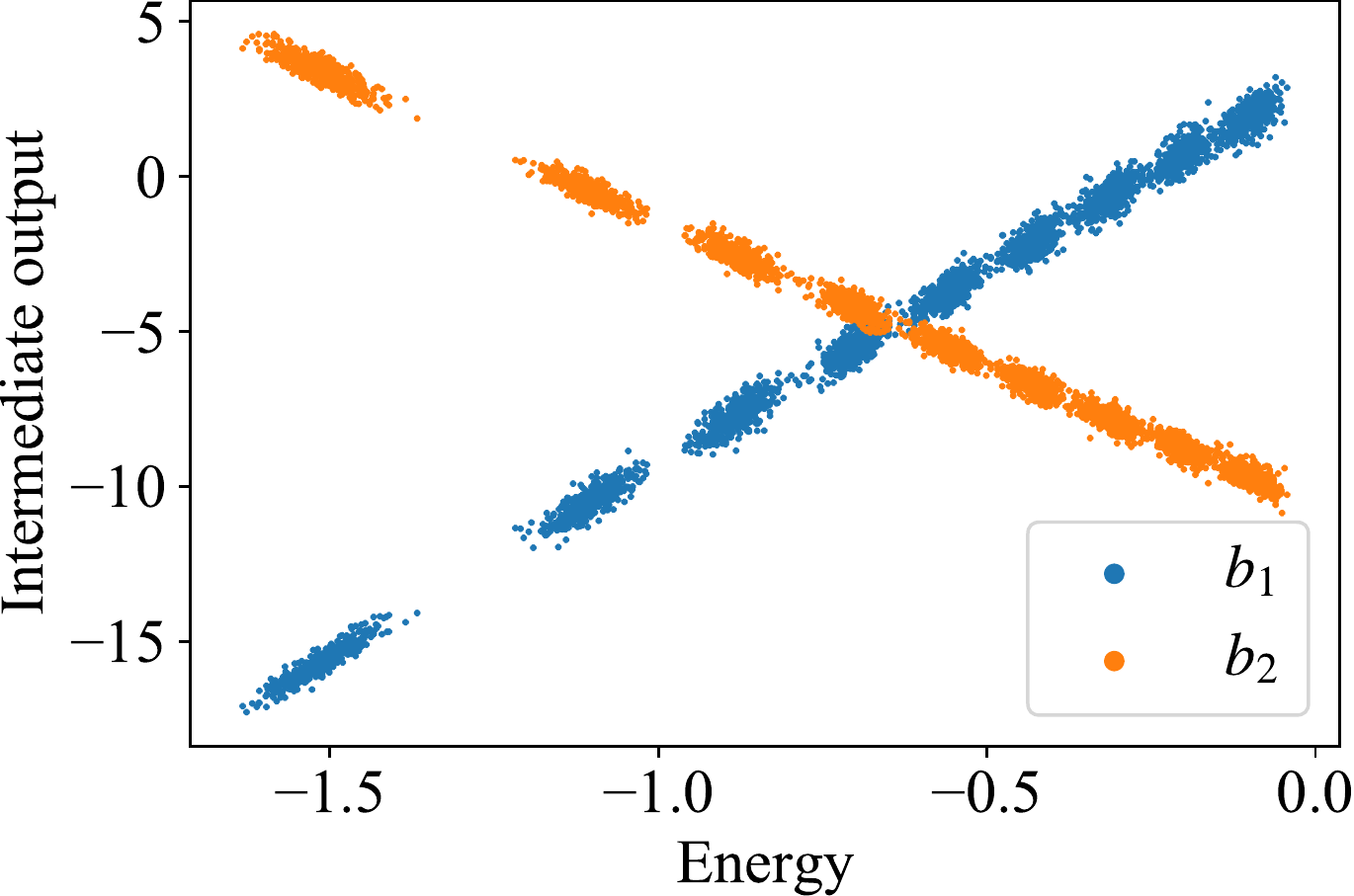}%
  \end{center}
  \caption{(Color online).
    Correlation between the intermediate output $\{ b_k \}$ and the energy in CNN trained by the double-random-gauge dataset.
    $b_1$ ($b_2$) shows a clear positive (negative) correlation.
  }
  \label{fig:G2-CNN-Out}
\end{figure}

\begin{figure}[tbp]
  \begin{center}
    \includegraphics[width=0.8\hsize]{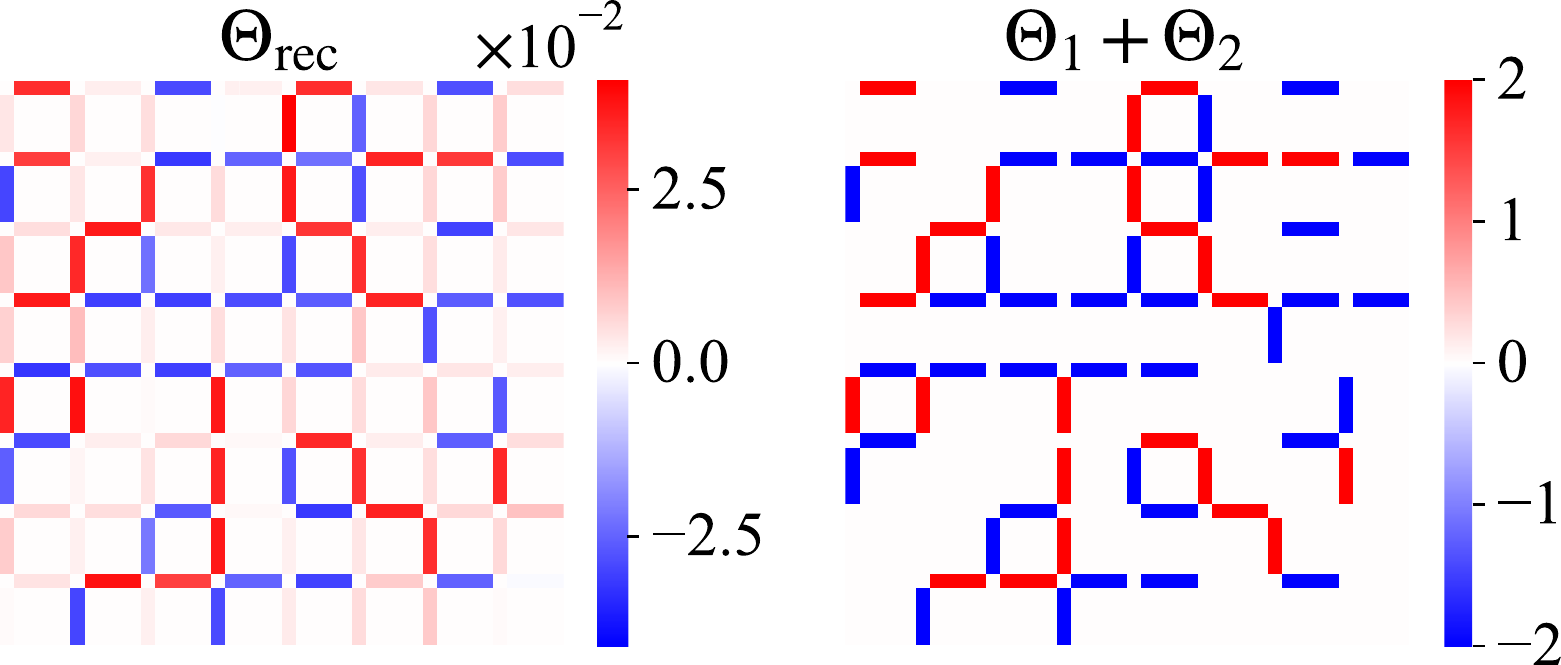}%
  \end{center}
  \caption{(Color online).
    Reconstructed gauge $\Theta_\mathrm{rec}$ from CNN trained by the double-random-gauge dataset.
    For comparison, we also show the sum of two gauges $\Theta_1+\Theta_2$.
    Only a part ($8 \times 8$) of the whole lattice ($64 \times 64$) is presented.
  }
  \label{fig:G2-CNN-Th}
\end{figure}

The mean classification accuracy is $89.4\%$, significantly lower than the single gauge case but still much higher than FNN.
The correlation between $\{b_k\}$ and the energy is shown in Fig.~\ref{fig:G2-CNN-Out}, in which one can see that CNN successfully extracts the energy from the double-random-gauge dataset.
We have tried to reconstruct the gauge $\Theta_\mathrm{rec}$ according to the same procedure proposed in Sec.~\ref{sec:Gauge-CNN} except that we did not apply the sign function for the final reconstructed gauge.
As shown in Fig.~\ref{fig:G2-CNN-Th}, we observed that the reconstructed gauge has a similar pattern as $\Theta _1 + \Theta _2$.
The mean correlation coefficient between the elements of $\Theta_\mathrm{rec}$ and $\Theta _1 + \Theta _2$ over 1000 trials is 0.918.
As in the FNN case, we observe two categories: the reconstructed gauges with correlation coefficient almost unity and those with small coefficient.
In the CNN case, the latter trials yield a correlation coefficient close to 0.7, where no filters have non-zero vertically (horizontally) aligned elements, as in the failure cases for the single gauge dataset.
On the other hand, in the former successful cases, we conclude that CNN detects lattice bonds where $\Theta_1$ and $\Theta_2$ have the same sign and uses them for extracting the local energy.

\section{Conclusion}
\label{sec:Conclusion}

In the present paper, we proposed a new test using the random-gauge Ising model for machine learning of the order parameter.
We have shown that although the order parameter is not directly visible in the random-gauge model, both FNN and CNN can extract the order parameter or the energy as in the ferromagnetic case.
We also discussed how and where the information of random gauge is encoded in NN and attempted to reconstruct the gauge from the NN parameters.
We found that FNN encodes the effect of random gauge to its weights naturally.
In contrast, CNN copes with the randomness by using only two elements among four in convolutional filters and assigning different network parts to local gauge patterns.
This observation indicates that although CNN demonstrated higher performance than FNN for the present randomized-gauge test, FNN is more efficient and suitable for dealing with models with spatial randomness.
In addition, the experiments for the double-random-gauge dataset revealed that NN could detect a part of lattice where two gauges have the same sign and use only that part for temperature prediction.
The present study demonstrated flexible and extraordinary NN ability in extracting the physical essence of cooperative phenomena from the data, which should help us design novel and robust machine learning algorithms for understanding many-body physics in the future.

\section*{Acknowlegements}

This work was partially supported by JSPS KAKENHI (No.~17K05564 and 20H01824).

\bibliography{main.bib}
\end{document}